\input harvmac
\input epsf
\baselineskip14pt
\newcount\figno
\figno=0
\def\fig#1#2#3{
\par\begingroup\parindent=0pt\leftskip=1cm\rightskip=1cm\parindent=0pt
\baselineskip=11pt
\global\advance\figno by 1
\midinsert
\epsfxsize=#3
\centerline{\epsfbox{#2}}
\vskip 12pt
{\bf Fig.\ \the\figno: } #1\par
\endinsert\endgroup\par
}
\def\figlabel#1{\xdef#1{\the\figno}}
\def\encadremath#1{\vbox{\hrule\hbox{\vrule\kern8pt\vbox{\kern8pt
\hbox{$\displaystyle #1$}\kern8pt}
\kern8pt\vrule}\hrule}}
\def\beq{\begin{equation}}
\def\eeq{\end{equation}}
\def\bea{\begin{eqnarray}}
\def\eea{\end{eqnarray}}
\def\zbar{\bar{z}}
\def\ubar{\bar{u}}
\def\alphab{\bar{\alpha}}
\def\betab{\bar{\beta}}
\def\Jbar{\bar{J}}
\def\Jtilde{\tilde{J}}

\def\bra{\langle}
\def\ket{\rangle}
\def\sa{\left[}
\def\sk{\right]}
\def\pat{\partial}
\def\patb{\bar{\partial}}
\def\gammab{\bar{\gamma}}
\def\Mbar{\bar{M}}
\def\mbar{\bar{m}}
\def\xbar{\bar{x}}
\def\ads{AdS$_3$}
\def\h3{H$_3$}
\def\half{{1 \over 2}}
\def\Dcal{{\cal D}}
\def\Ccal{{\cal C}}


\Title{\vbox{\baselineskip12pt
\hbox{hep-th/0208003}
\hbox{HIP-2002-30/TH}
\hbox{UCLA-02-TEP-21}
\vskip-.5in}}
{\vbox{\centerline{Strings in the Extended BTZ Spacetime }}}

\medskip\bigskip
\centerline{Samuli Hemming$^1$, Esko Keski-Vakkuri$^1$, and Per Kraus$^{2}$}
\bigskip\medskip
\centerline{\it $^1$Helsinki Institute of Physics,}
\centerline{\it  P.O. Box 64,
FIN-00014 University of Helsinki, Finland}
\centerline{\tt samuli.hemming@helsinki.fi, esko.keski-vakkuri@helsinki.fi}
\medskip
\centerline{\it $^2$Department of Physics and Astronomy}
\centerline{\it UCLA, Los Angleles, CA 90095-1547, U.S.A.}
\centerline{\tt pkraus@physics.ucla.edu}
\medskip\bigskip\medskip\bigskip\medskip
\baselineskip14pt

\noindent
We study string theory on the extended spacetime of the
BTZ black hole, as described by an orbifold of the SL(2,R)
WZW model.  The full spacetime has an infinite number of disconnected
boundary components, each corresponding to a dual CFT.  We discuss
the computation of  bulk and boundary correlation functions for
operators inserted on different components.  String theory correlation
functions are obtained by analytic continuation from an orbifold of
the SL(2,C)/SU(2)  coset model.  This yields two-point functions for
general operators,   including those describing strings that wind
around the horizon  of the black hole.

\Date{July, 2002}

\lref\MOI{
J.~M.~Maldacena and H.~Ooguri,
J.\ Math.\ Phys.\  {\bf 42}, 2929 (2001)
[arXiv:hep-th/0001053].
}

\lref\HenningsonJC{ M.~Henningson, S.~Hwang, P.~Roberts and
B.~Sundborg,
Phys.\ Lett.\ B {\bf 267}, 350 (1991).
}

\lref\HwangUK{ S.~Hwang and P.~Roberts,
arXiv:hep-th/9211075.
}

\lref\NatsuumeIJ{ M.~Natsuume and Y.~Satoh,
Int.\ J.\ Mod.\ Phys.\ A {\bf 13}, 1229 (1998)
[arXiv:hep-th/9611041].
}

\lref\MartinecCF{
E.~J.~Martinec and W.~McElgin,
JHEP {\bf 0204}, 029 (2002)
[arXiv:hep-th/0106171].
}

\lref\SatohXE{ Y.~Satoh,
Nucl.\ Phys.\ B {\bf 513}, 213 (1998) [arXiv:hep-th/9705208].
}

\lref\MOS{
J.~M.~Maldacena, H.~Ooguri and J.~Son,
J.\ Math.\ Phys.\  {\bf 42}, 2961 (2001)
[arXiv:hep-th/0005183].
}

\lref\KutasovXU{
D.~Kutasov and N.~Seiberg,
JHEP {\bf 9904}, 008 (1999)
[arXiv:hep-th/9903219].
}

\lref\deBoerPP{
J.~de Boer, H.~Ooguri, H.~Robins and J.~Tannenhauser,
JHEP {\bf 9812}, 026 (1998)
[arXiv:hep-th/9812046].
}

\lref\MOIII{
J.~M.~Maldacena and H.~Ooguri,
Phys.\ Rev.\ D {\bf 65}, 106006 (2002)
[arXiv:hep-th/0111180].
}

\lref\CruzIR{
N.~Cruz, C.~Martinez and L.~Pena,
Class.\ Quant.\ Grav.\  {\bf 11}, 2731 (1994)
[arXiv:gr-qc/9401025].
}

\lref\MartinecXQ{
E.~J.~Martinec and W.~McElgin,
arXiv:hep-th/0206175.
}

\lref\GiveonNS{
A.~Giveon, D.~Kutasov and N.~Seiberg,
Adv.\ Theor.\ Math.\ Phys.\  {\bf 2}, 733 (1998)
[arXiv:hep-th/9806194].
}

\lref\SonQM{
J.~Son,
arXiv:hep-th/0107131.
}

\lref\TroostWK{
J.~Troost,
arXiv:hep-th/0206118.
}

\lref\HananyEV{
A.~Hanany, N.~Prezas and J.~Troost,
JHEP {\bf 0204}, 014 (2002)
[arXiv:hep-th/0202129].
}

\lref\DanielssonZT{
U.~H.~Danielsson, E.~Keski-Vakkuri and M.~Kruczenski,
Nucl.\ Phys.\ B {\bf 563}, 279 (1999)
[arXiv:hep-th/9905227].
}
\lref\EvansFR{
T.~S.~Evans, A.~Gomez Nicola, R.~J.~Rivers and D.~A.~Steer,
arXiv:hep-th/0204166.
}

\lref\BanadosWN{
M.~Banados, C.~Teitelboim and J.~Zanelli,
Phys.\ Rev.\ Lett.\  {\bf 69}, 1849 (1992)
[arXiv:hep-th/9204099];
M.~Banados, M.~Henneaux, C.~Teitelboim and J.~Zanelli,
Phys.\ Rev.\ D {\bf 48}, 1506 (1993)
[arXiv:gr-qc/9302012].}

\lref\ElitzurRT{
S.~Elitzur, A.~Giveon, D.~Kutasov and E.~Rabinovici,
JHEP {\bf 0206}, 017 (2002)
[arXiv:hep-th/0204189].
}

\lref\CrapsII{
B.~Craps, D.~Kutasov and G.~Rajesh,
JHEP {\bf 0206}, 053 (2002)
[arXiv:hep-th/0205101].
}

\lref\CornalbaNV{
L.~Cornalba, M.~S.~Costa and C.~Kounnas,
arXiv:hep-th/0204261.
}

\lref\GubserBC{
S.~S.~Gubser, I.~R.~Klebanov and A.~M.~Polyakov,
Phys.\ Lett.\ B {\bf 428}, 105 (1998)
[arXiv:hep-th/9802109].
}

\lref\KeskiVakkuriNW{
E.~Keski-Vakkuri,
Phys.\ Rev.\ D {\bf 59}, 104001 (1999)
[arXiv:hep-th/9808037].
}

\lref\HemmingWE{S.~Hemming and E.~Keski-Vakkuri,
Nucl.\ Phys.\ B {\bf 626}, 363 (2002) [arXiv:hep-th/0110252].
}

\lref\LifschytzEB{
G.~Lifschytz and M.~Ortiz,
Phys.\ Rev.\ D {\bf 49}, 1929 (1994)
[arXiv:gr-qc/9310008].
}

\lref\BalasubramanianSN{
V.~Balasubramanian, P.~Kraus and A.~E.~Lawrence,
Phys.\ Rev.\ D {\bf 59}, 046003 (1999)
[arXiv:hep-th/9805171].
}

\lref\WittenQJ{
E.~Witten,
Adv.\ Theor.\ Math.\ Phys.\  {\bf 2}, 253 (1998)
[arXiv:hep-th/9802150].
}

\lref\BalasubramanianRE{
V.~Balasubramanian and P.~Kraus,
Commun.\ Math.\ Phys.\  {\bf 208}, 413 (1999)
[arXiv:hep-th/9902121].
}

\lref\HorowitzXK{
G.~T.~Horowitz and D.~Marolf,
JHEP {\bf 9807}, 014 (1998)
[arXiv:hep-th/9805207].
}

\lref\BalasubramanianDE{
V.~Balasubramanian, P.~Kraus, A.~E.~Lawrence and S.~P.~Trivedi,
Phys.\ Rev.\ D {\bf 59}, 104021 (1999)
[arXiv:hep-th/9808017].
}

\lref\CarneirodaCunhaNW{
B.~G.~Carneiro da Cunha,
Phys.\ Rev.\ D {\bf 65}, 104025 (2002)
[arXiv:hep-th/0110169].
}

\lref\MaldacenaBW{
J.~M.~Maldacena and A.~Strominger,
JHEP {\bf 9812}, 005 (1998)
[arXiv:hep-th/9804085].
}

\lref\MaldacenaKR{
J.~M.~Maldacena,
arXiv:hep-th/0106112.
}

\lref\GiveonNS{
A.~Giveon, D.~Kutasov and N.~Seiberg,
Adv.\ Theor.\ Math.\ Phys.\  {\bf 2}, 733 (1998)
[arXiv:hep-th/9806194].
}

\lref\TeschnerFT{
J.~Teschner,
Nucl.\ Phys.\ B {\bf 546}, 390 (1999)
[arXiv:hep-th/9712256];
Nucl.\ Phys.\ B {\bf 571}, 555 (2000)
[arXiv:hep-th/9906215].
}

\lref\HorowitzJC{
G.~T.~Horowitz and D.~L.~Welch,
Phys.\ Rev.\ Lett.\  {\bf 71}, 328 (1993)
[arXiv:hep-th/9302126].
}

\lref\KaloperKJ{
N.~Kaloper,
Phys.\ Rev.\ D {\bf 48}, 2598 (1993)
[arXiv:hep-th/9303007].
}

\lref\BalasubramanianRY{
V.~Balasubramanian, S.~F.~Hassan, E.~Keski-Vakkuri and A.~Naqvi,
arXiv:hep-th/0202187.
}

\lref\CornalbaFI{
L.~Cornalba and M.~S.~Costa,
arXiv:hep-th/0203031.
}

\lref\NekrasovKF{
N.~A.~Nekrasov,
arXiv:hep-th/0203112.
}

\lref\SimonMA{
J.~Simon,
JHEP {\bf 0206}, 001 (2002)
[arXiv:hep-th/0203201].
}

\lref\LiuFT{
H.~Liu, G.~Moore and N.~Seiberg,
JHEP {\bf 0206}, 045 (2002)
[arXiv:hep-th/0204168];
arXiv:hep-th/0206182.
}

\lref\LawrenceAJ{
A.~Lawrence,
arXiv:hep-th/0205288.
}

\lref\FabingerKR{
M.~Fabinger and J.~McGreevy,
arXiv:hep-th/0206196.
}

\lref\HorowitzMW{
G.~T.~Horowitz and J.~Polchinski,
arXiv:hep-th/0206228.
}

\lref\SusskindQC{
L.~Susskind and J.~Uglum,
Nucl.\ Phys.\ Proc.\ Suppl.\  {\bf 45BC}, 115 (1996)
[arXiv:hep-th/9511227].
}

\newsec{Introduction}

The BTZ black hole spacetime  \BanadosWN\ possesses
many features that one would
like to understand better in
string theory:  event horizons, Hawking radiation, time dependence,
nontrivial causal structure with
potential closed timelike curves, etc.   Since the corresponding
worldsheet theory  is an orbifold of the SL(2,R) WZW model, classical
string theory is in principle exactly solvable in this background.
Furthermore, being asymptotically \ads, the theory has a dual
holographic description as a $1+1$ dimensional CFT.   For all these
reasons, it seems fruitful to gain a detailed understanding of string
theory in the BTZ spacetime.

Early work concerning strings on BTZ includes
\refs{\HorowitzJC,\KaloperKJ,\NatsuumeIJ,\SatohXE}.   However,
progress was delayed by an incomplete understanding of the
underlying SL(2,R) WZW model: an {\it ad hoc} cutoff on the
spectrum seemed to be needed for unitarity.   It is now known
\MOI\ that instead of imposing a cutoff one should include long
strings and spectral flowed states
in the spectrum (see also \refs{\HenningsonJC,\HwangUK}), and that
the resulting theory is unitary. In light of this new
understanding, \HemmingWE\ elaborated on the earlier work
\NatsuumeIJ\ on the string spectrum in BTZ and interpreted it in
the context of spectral flow. Additional related work can
be found in \refs{\GiveonNS,\MartinecCF,
\SonQM,\MartinecXQ,\TroostWK}. Here we would like to continue this
program, focussing on the string theory interpretation of the
extended BTZ geometry, and evaluating some simple correlation
functions in this background.

The maximal extension of the rotating BTZ black hole has an intricate
causal structure with multiple asymptotic regions, analogous to the
Kerr solution in asymptotically flat spacetime.    The multiple boundaries
of the spacetime lead to a richer example of holography than usual,
with the possibility of computing correlation functions of operators
inserted on disconnected boundary components. This is similar to what
one can expect for certain cosmological spacetimes, with distinct boundaries
in the far past and future; for recent examples see
\refs{\ElitzurRT,\CrapsII,\CornalbaNV}.   We would like
to know the rules for relating bulk and boundary correlation functions in
such a situation.\foot{The physical relevance of the extended spacetime
can be questioned due to potentially destabilizing backreaction effects;
we discuss this more in the text.}

Given a spacetime with spacelike separated disconnected boundaries,
it is known that the Hilbert space of the dual CFT is the product
of the CFT Hilbert spaces corresponding to the distinct boundary
components
\refs{\HorowitzXK,\BalasubramanianDE,\CarneirodaCunhaNW,\MaldacenaKR}.
In the case of a black hole, tracing over an unobserved
Hilbert space yields a thermal density matrix for the remaining space.
As discussed in \MaldacenaKR\ this can be understood by analytic
continuation from the Euclidean black hole; for example wavefunctions
in the left and right halves of the Kruskal diagram for a nonrotating
black hole are related by the imaginary time evolution $t \rightarrow
t + i\beta/2$, yielding a Boltzmann factor.
   In the case of string theory, analytic continuation
from Euclidean signature takes on added significance, since we do not
at present know how to compute correlation functions directly in
Lorentzian signature.
As we discuss, correlations among non-spacelike separated boundary
components can also be found by continuation from Euclidean signature,
and the result can again be related to correlation functions in a
tensor product Hilbert space.  This gives a holographic interpretation
of the extended BTZ solution.

String theory correlation functions in  \ads\ were obtained in \MOIII\
by analytically continuing results from the
SL(2,C)/SU(2) model \TeschnerFT .  We can apply
the same strategy in the BTZ case, starting from the appropriate
orbifold of the SL(2,C)/SU(2) model.   The main difference is that
one needs to work in the hyperbolic basis for the current algebra,
rather than the elliptic basis normally used for \ads.   In this basis,
the spectral flow operation of \MOI\  generates strings that wind around
the  black hole horizon \HemmingWE.   We will focus on the two point
functions for
vertex operators of flowed and unflowed string states. Ultimately, one
would like to describe interaction in this background, including loop
effects, in order to see what string theory has to say about the BTZ
singularity.

The remainder of this paper is organized as follows.  In section 2 we review
the geometry of the extended BTZ spacetime, in particular the structure
of the boundary, and the relations between the different coordinate patches.
Section 3 discusses correlation functions in the field theory limit.
We review some relevant aspects of \ads\ string theory in section 4.
String theory correlation functions in BTZ are computed in section 5,
and in section 6 we conclude with a discussion of some open problems.
In an attempt  to make this paper approximately self-contained we have
included a substantial amount of review material.  A reader who is
very familiar
with the extended BTZ geometry can safely skim much of section 2, and similarly
section 4 for the reader well versed in \ads\ string theory.

\newsec{Causal structure of the BTZ black hole}

We begin by reviewing the extended BTZ geometry, paying special attention
to the structure of the boundary.   The BTZ geometry is of course well
understood from the orginal work \BanadosWN.  In those papers it was
proposed to truncate the geometry at a ``singularity''
in order to avoid the presence of closed timelike curves.  Ultimately,
this can only be justified by doing calculations in the full quantum theory,
since closed timelike curves can be consistent at the classical level.
Here our focus is on classical string physics and so we will keep the full
spacetime including the regions with closed timelike curves.

\subsec{Lorentzian black hole}

The BTZ black hole is obtained by making identifications in \ads.
AdS$_3$ is defined by the hyperboloid
\eqn\aa{x_0^2 + x_1^2 - x_2^2 -x_3^2 = \ell^2.}
This is also the SL(2,R) group manifold,
\eqn\ab{g = {1 \over \ell}\pmatrix{ x_1 + x_2 & x_3 +x_0 \cr
x_3-x_0 & x_1 -x_2}, \quad  \det g = 1.}
Henceforth, we will always consider the covering space of the group
manifold, sometimes denoted as CAdS$_3$.
The AdS$_3$ metric is the invariant metric on the group manifold,
\eqn\ab{ ds^2 =-\ell^2 {\rm Tr} g^{-1} dg\, g^{-1} dg.}
The isometry group of AdS$_3$ is therefore SL(2,R)$_L$ $\times$ SL(2,R)$_R$,
acting as
\eqn\ac{ g ~ \rightarrow ~ g_L g g_R.}
SL(2,R) has three types of conjugacy classes:
\eqn\ad{\eqalign{
{\rm hyperbolic:}\quad &|{\rm Tr}\, g | > 2, \quad
g = h \pmatrix{\alpha & 0 \cr 0 & \alpha^{-1}}h^{-1}, \cr
{\rm elliptic:}\quad &|{\rm Tr}\, g | < 2, \quad
g = h \pmatrix{\cos \alpha & \sin \alpha \cr -\sin \alpha
 & \cos \alpha}h^{-1}, \cr
{\rm parabolic:}\quad &|{\rm Tr}\, g | = 2, \quad
g = \pm h \pmatrix{1 & 1\cr 0 & 1}h^{-1}. }}
To define the BTZ black hole we identify by elements of a hyperbolic
conjugacy class
\eqn\ae{ g \cong \rho_L g \rho_R,}
with
\eqn\af{\eqalign{ \rho_L & = \pmatrix{ e^{2\pi^2 T_+} & 0 \cr 0 &
e^{-2\pi^2 T_+}}, \cr & \cr
\rho_R & = \pmatrix{ e^{2\pi^2 T_-}  & 0 \cr 0 &
e^{-2\pi^2 T_-} },}}
and $T_+ \leq T_-$.
The radii of the inner and outer horizons of the black hole,
$r_-$ and $r_+$, are related to $T_\pm$ by
\eqn\aea{\eqalign{r_+ & = \pi \ell(T_+ + T_-),\cr
 \quad r_-& =-\pi \ell(T_+-T_-).}}
 The mass and angular momentum of the black hole are
\eqn\ag{\eqalign{M & =2\pi^2(T_+^2 +T_-^2) = {r_+^2 + r_-^2 \over \ell^2} \cr
J & = -2\pi^2 \ell(T_+^2 -T_-^2)={2 r_+ r_-\over \ell}.}}
The identifications \af\ act as a boost in the $x_1 - x_2$ and $x_0 - x_3$ planes:
\eqn\aha{ \pmatrix{ x_1 \cr x_2 \cr x_3 \cr x_0} ~ \cong ~ \pmatrix{
\cosh \gamma_+ & \sinh \gamma_+ & 0 &0 \cr
\sinh \gamma_+ & \cosh \gamma_+&0&0\cr
0&0&\cosh \gamma_-&-\sinh \gamma_-\cr 0&0&-\sinh \gamma_-&\cosh \gamma_-}
\pmatrix{ x_1 \cr x_2\cr x_3 \cr x_0},
\quad \gamma_{\pm} = {2\pi r_\pm \over \ell}.}
The identification has no fixed points in the rotating case with $r_- \neq 0$.
In the non-rotating case $r_- = 0$ there are fixed points at $x_1 = x_2 =0$.
This line of fixed points is what BTZ call the singularity
of the non-rotating BTZ black hole.

The identification \aha\ gives rise to closed timelike curves.  In
the non-rotating case this
is  easily seen by examining the geometry in the neighborhood of the
 fixed points,
\eqn\ai{\eqalign{ x_1 & = {\rm O}(\epsilon), \cr
x_2 & = {\rm O}(\epsilon), \cr
x_3 & = \pm \sqrt{- \ell^2 + x_0^2} + {\rm O}(\epsilon^2)}}
giving the metric
\eqn\aj{ ds^2 =  -dx_1^2 + dx_2^2 + { dx_0^2 \over
1 - x_0^2/\ell^2}  + {\rm O}(\epsilon^2).}
So near the fixed point the identification \aha\ acts as a boost in $R^{1,1}$,
which is a timelike identification for $|x_2| > |x_1|$.
In the rotating case the identification \aha\ also shifts $x_0$, thereby
smoothing out the orbifold.

To get a picture of the global structure we divide the original
AdS$_3$ manifold into the following 3 types of regions
\eqn\ak{\eqalign{{\rm Region~ 1:}\quad & x_1^2 - x_2^2 \geq 0, \quad
x_0^2 - x_3^2 \leq 0, \cr
{\rm Region~ 2:}\quad & x_1^2 - x_2^2 \geq 0, \quad
x_0^2 - x_3^2 \geq 0, \cr
{\rm Region~ 3:}\quad & x_1^2 - x_2^2 \leq 0, \quad
x_0^2 - x_3^2 \geq 0.}}

The original AdS$_3$ manifold is best thought of as a solid cylinder with an
$S^1 \times R$ boundary.   Which regions reach the boundary?  The
AdS$_3$ boundary is given by taking (this will become more apparent
when we introduce explicit coordinates)
\eqn\al{\eqalign{|x_1^2 - x_2^2| ~ & \rightarrow ~ \infty \cr
|x_0^2 - x_3^2| ~ & \rightarrow ~ \infty,}}
subject of course to \aa.   From \ak\ and \al\ it is apparent that
region 2 does not extend out to the boundary, while regions 1 and 3 do.
Regions 1 and 2 are both bounded by an  asymptotic boundary at infinity
and an event horizon.  So from the perspective of either region 1 or 3
 the other two regions lie behind the horizon.  As we'll review, region 3
 contains closed timelike curves,  and the ``singularity'' of the
BTZ black hole is located at the boundary of the
ergosphere ($g_{tt} =0$ in stationary
coordinates) of region 3.

The geometry is easier to visualise in the non-rotating case, so we
consider this first.
The boundaries separating adjacent regions define null hypersurfaces
in AdS$_3$; drawing these in the \ads\ cylinder we obtain figure 1.
\fig{The AdS$_3$ cylinder. Depicted in the figure are the null hypersurfaces
separating regions 2 and 3, as well as two colored
 vertical cross sections yielding Penrose diagrams. The
standard Penrose diagram appears as the red square, also displayed
in figure 2. The perpendicular blue square gives the Penrose
diagram of figure 3.}{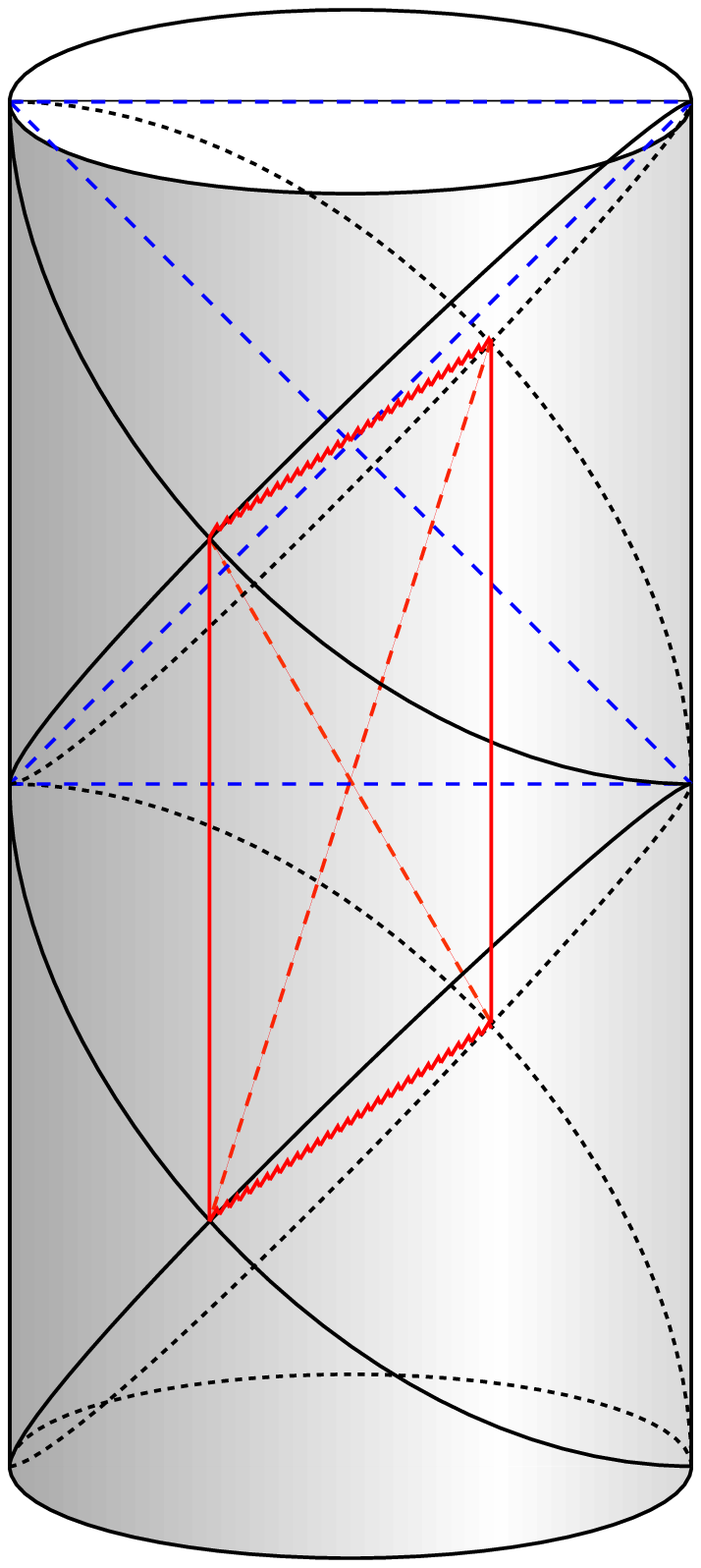}{2.0truein}
Penrose diagrams
are obtained by drawing two  dimensional vertical cross sections of the
cylinder, as in figures 2 and 3.
\fig{The standard non-rotating BTZ Penrose diagram.}{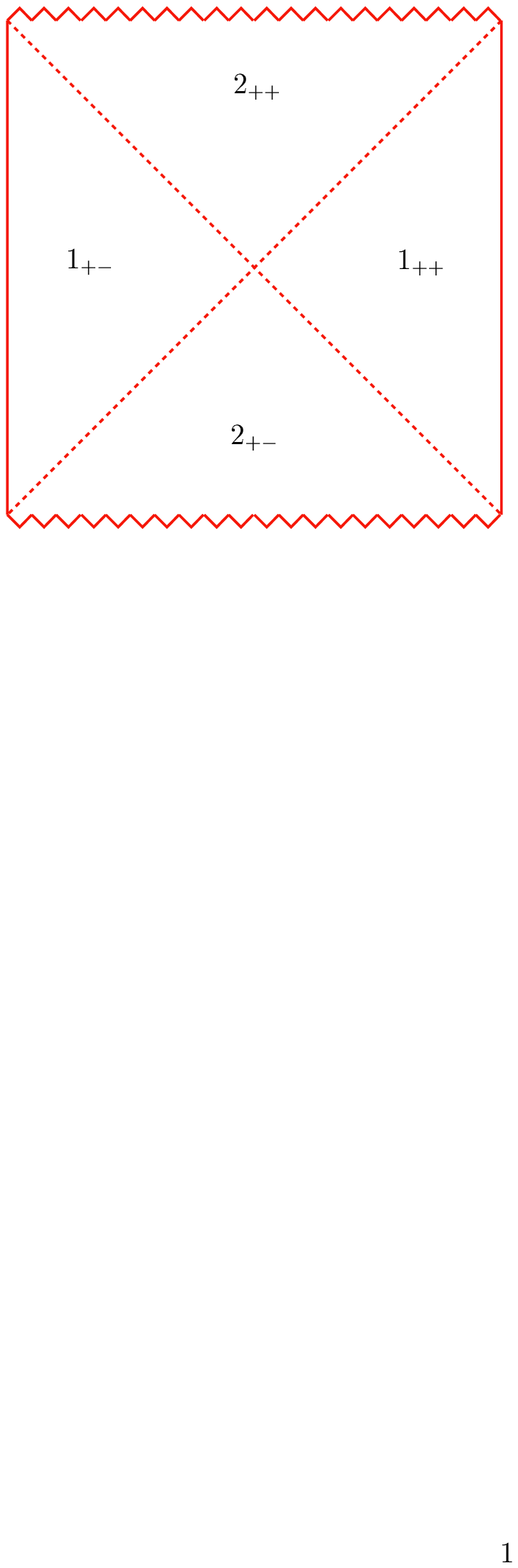}{2.0truein}
\fig{The Penrose diagram showing region 3, with
identifications indicated}{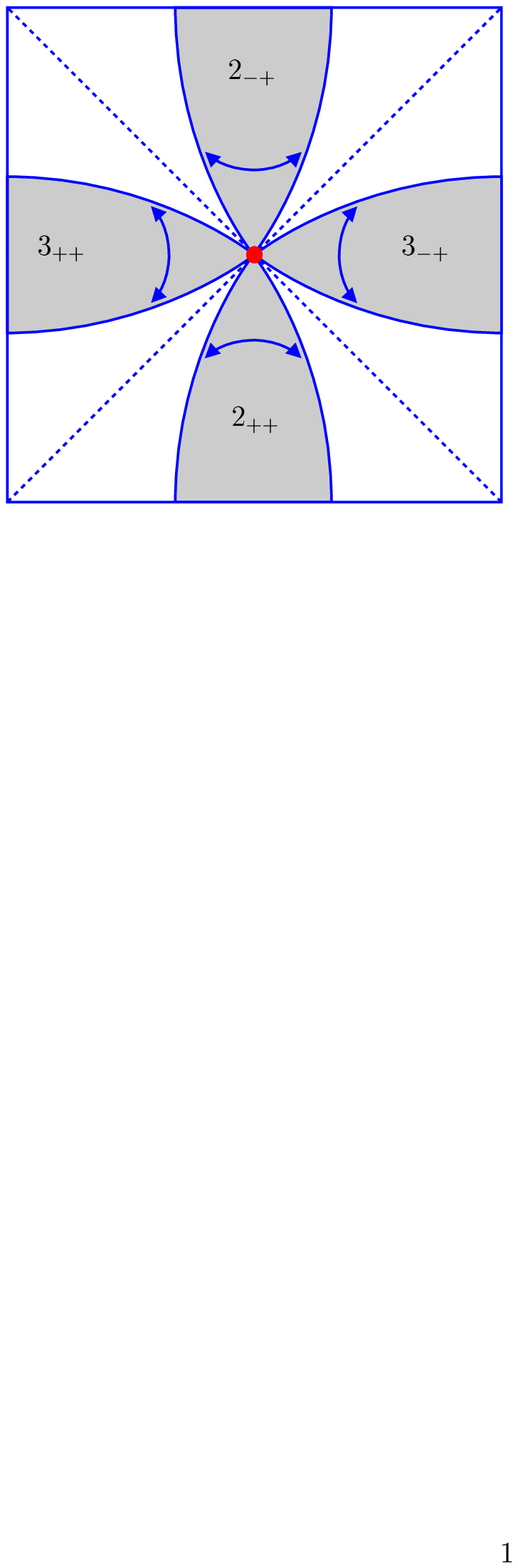}{2.0truein}
The  Penrose diagram in figure 2 is the
standard one with the singularity appearing as a spacelike hypersurface,
and the region 3 containing the closed timelike curves
does not appear.  In figure 1, which depicts the AdS cylinder, the standard
diagram appears as the rectangle in the center\foot{Shown in red if colors are
displayed.}. The perpendicular vertical
cross section of the cylinder\foot{Shown in blue.} gives
the  Penrose diagram of figure 3. In the latter
Penrose diagram region 3 is displayed while region 1 is absent.  See
\MartinecXQ\ for some other depictions of the geometry.

We now  introduce explicit  coordinates for the general rotating black hole.
The BTZ identifications will preserve two Killing vectors out
of the original six, and we take $u_+$ and $u_-$ to be coordinates labelling
the orbits of these Killing vectors, as well as the radial coordinate $r$.
  This requires that we cover each
of the regions 1,2,3 by four separate coordinate patches.
We henceforth
work in units where $\ell =1$.  In the following $\eta_{1,2} = \pm 1$.

\noindent
{\bf Region 1:}
\eqn\am{\eqalign{
x_1 & = \eta_1\left( {r^2 - r_-^2 \over r_+^2 - r_-^2}\right)^{1/2}
\cosh \pi(T_+ u_+ +T_- u_-) \cr
x_2 & = \eta_1 \left( {r^2 - r_-^2 \over r_+^2 - r_-^2}\right)^{1/2}
\sinh \pi(T_+ u_+ +T_- u_-) \cr
x_3 & = \eta_2 \left({r^2 - r_+^2 \over r_+^2 - r_-^2}\right)^{1/2}
 \cosh\pi(T_+ u_+ - T_- u_-) \cr
x_0 & = \eta_2 \left({r^2 - r_+^2 \over r_+^2 - r_-^2}\right)^{1/2}
  \sinh\pi(T_+ u_+ - T_- u_-).
}}

\noindent
{\bf Region 2:}
\eqn\an{\eqalign{
x_1 & = \eta_1\left( {r^2 - r_-^2 \over r_+^2 - r_-^2}\right)^{1/2} \cosh \pi(T_+ u_+ +T_- u_-) \cr
x_2 & = \eta_1 \left( {r^2 - r_-^2 \over r_+^2 - r_-^2}\right)^{1/2} \sinh \pi(T_+ u_+ +T_- u_-) \cr
x_3 & =  \eta_2 \left({r_+^2 - r^2 \over r_+^2 - r_-^2}\right)^{1/2}
  \sinh\pi(T_+ u_+ - T_- u_-) \cr
x_0 & = \eta_2 \left({r_+^2 - r^2 \over r_+^2 - r_-^2}\right)^{1/2}
 \cosh\pi(T_+ u_+ - T_- u_-).
}}

\noindent
{\bf Region 3:}
\eqn\ao{\eqalign{
x_1 & =  \eta_1 \left({r^2 - r_+^2 \over r_+^2 - r_-^2}\right)^{1/2}
  \sinh\pi(T_+ u_+ - T_- u_-) \cr
x_2 & =\eta_1 \left({r^2 - r_+^2 \over r_+^2 - r_-^2}\right)^{1/2}
 \cosh\pi(T_+ u_+ - T_- u_-) \cr
x_3 & = \eta_2 \left( {r^2 - r_-^2 \over r_+^2 - r_-^2}\right)^{1/2} \sinh \pi(T_+ u_+ +T_- u_-) \cr
x_0 & = \eta_2 \left( {r^2 - r_-^2 \over r_+^2 - r_-^2}\right)^{1/2} \cosh \pi(T_+ u_+ +T_- u_-).
}}
In all three regions $u_\pm$  range over all real values.
$r_+ \leq r \leq \infty$ in regions 1 and 3;
$r_- \leq r \leq r_+$ in region 2.

We define $t$ and $\phi$ by
\eqn\asa{u_\pm = \phi \pm t.}
The BTZ identification  \ae\ in these coordinates is
\eqn\as{\eqalign{{\rm Regions ~1,2:}\quad &
(t,\phi,r) ~ \cong ~ (t,\phi+2\pi,r) \cr
{\rm Region ~3:}\quad & (t,\phi,r) ~ \cong ~ (t+2\pi,\phi,r).
}}
The solution written in $t,\phi,r$ coordinates is
\eqn\at{\eqalign{ ds^2 & = -{ (r^2 -r_+^2)(r^2-r_-^2) \over r^2}dt^2
+{r^2 \over (r^2 -r_+^2)(r^2 -r_-^2)}dr^2
+r^2(d\phi - {r_+ r_- \over r^2} dt)^2.}}
In string theory we also have a nonvanishing NS-NS B-field.  In these
coordinates it is
\eqn\asa{\eqalign{
B & = \left\{
\eqalign{ &(r^2- r_-^2) d\phi \wedge dt \quad{\rm regions~1,2} \cr
& (r^2- r_+^2) d\phi \wedge dt \quad{\rm region~3.}} \right.
}}
More precisely, we have twelve patches corresponding to
regions 1,2,3 and four choices
for $\eta_{1,2}$.  The full spacetime consists of an infinite
vertical stack of these patches.  The Penrose diagram of the rotating
black hole is obtained from \at.  Dropping the third term in \at,
writing the remainder as $ds^2 = \Omega^2(x^+,x^-)dx^+ dx^-$, and assembling
the different patches, we arrive at figure 4.  This figure indicates the
causal structure in the $t-r$ plane.  However, note that null geodesics
will not remain in this plane; this in contrast to the Penrose diagram
for the four dimensional Kerr solution, which is drawn along the axis of
symmetry of the black hole.
\fig{The Penrose diagram of the rotating BTZ black hole. Dashed lines
represent what BTZ called the singularity.
 Arrows indicate the identifications, which
become timelike when extended past the singularity.}{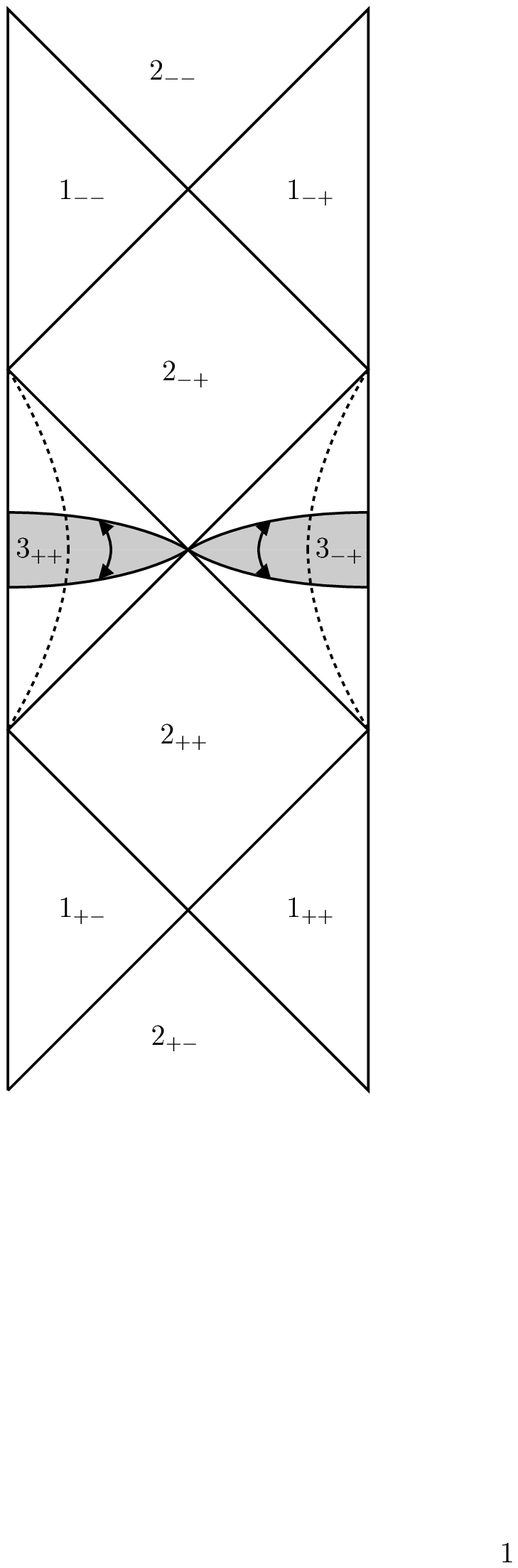}{2.0truein}
We chose our coordinates
so that the metric takes the same form in all three regions and so that $t$ is
a timelike coordinate for large $r$, but note
that this implies that $r$ jumps when we cross from region 2 to 3.  In
particular, the boundary between regions 2 and 3 is at $r=r_-$ when viewed
from region 2, and $r=r_+$ when viewed from region 3.  For the same reason
the B-field changes form in \asa,
though the change is just a gauge transformation.

From \at\ it is clear that the BTZ identification is spacelike in
regions 1 and
2, and timelike in region 3 for $r > (r_-^2 + r_+^2)^{1/2}$. What is referred
to as the singularity of the rotating BTZ solution is the boundary of
the ergosphere,
$r=(r_-^2 + r_+^2)^{1/2}$ in region 3.  If the geometry is truncated here
then there will be no closed timelike curves.

For computing correlation functions in the extended BTZ geometry it
is very useful to note that we can use analytic continuation to go
from one region to another.  For instance, starting in region $1_{++}$
(meaning $\eta_1=\eta_2 = +1$) we can continue to the other three regions
$1_{\eta_1 \eta_2}$ by making the replacements
\eqn\ava{\eqalign{ 1_{+-}: \quad T_{\pm} u_\pm ~ & \rightarrow  ~
T_{\pm}u_\pm \mp {i \over 2} \cr
1_{-+}: \quad T_{\pm} u_\pm ~ & \rightarrow  ~
 T_{\pm}u_\pm -{i \over 2} \cr
1_{--}: \quad T_+ u_+ ~ & \rightarrow  ~ T_+ u_+ -i \cr
u_- ~ & \rightarrow ~ u_-.}}
Similarly, to go from region $1_{\eta_1 \eta_2}$ to region
$3_{\eta_1 \eta_2}$   we take
\eqn\ax{\eqalign{  T_+ u_+  ~ & \rightarrow ~ T_+ u_+ - {i \over 2} \cr
 u_- ~ & \rightarrow ~ -u_- \cr
r^2 ~ & \rightarrow ~   r_+^2 + r_-^2-r^2.}}
We chose the signs of the imaginary parts for later convenience; flipping
these just takes one to another copy of the respective region.

\subsec{Boundary structure}

Regions 1 and 3 have boundaries at $r \rightarrow \infty$.  The metric
on the boundary is
\eqn\ba{ds^2 = r^{2}(-dt^2 + d\phi^2)}
with the identifications \as.   In both cases the boundary is an
 infinite cylinder,
with the circle direction being spacelike in region 1 and timelike in
region 3.

It is helpful to study how the boundary of the original AdS$_3$ cylinder
is broken up by the identifications.  Global coordinates for AdS$_3$ are
\eqn\ca{\eqalign{ x_1 & = \cosh \mu \cos \tau \cr
x_2 &= \sinh \mu \sin \theta  \cr
x_3 & = \sinh \mu \cos \theta \cr
x_0 & = \cosh \mu \sin \tau,
}}
with metric
\eqn\cb{ds^2 = -\cosh^2 \mu\, d\tau^2 + d\mu^2 + \sinh^2 \mu\, d\theta^2.
}
Consider the boundary region, $\mu \rightarrow \infty$.   The boundary is
conformal to the cylinder $ds^2 = - d\tau^2 + d\theta^2$, with
$\theta \cong \theta + 2\pi$.   At the boundary the coordinate transformation
between global and BTZ coordinates becomes

\noindent
{\bf Region 1:}
\eqn\cc{\tan{\tau \pm \theta \over 2}
= \left(\pm \tanh \pi T_\pm u_\pm\right)^{\eta_1 \eta_2}}
\noindent
{\bf Region 3:}
\eqn\cc{\tan{\tau \pm \theta \over 2}  =
\left(\pm \tanh \pi T_\pm u_\pm\right)^{-\eta_1 \eta_2}.}
The original boundary in global coordinates therefore breaks up into
eight separate patches, repeated with periodicity $\Delta \tau = 2\pi$;
see Figure 5.
\fig{The boundary. Shaded areas are fundamental domains under
the identifications. Lines with double arrows indicate the identifications.
Lines with single arrows indicate
the flow of time.}{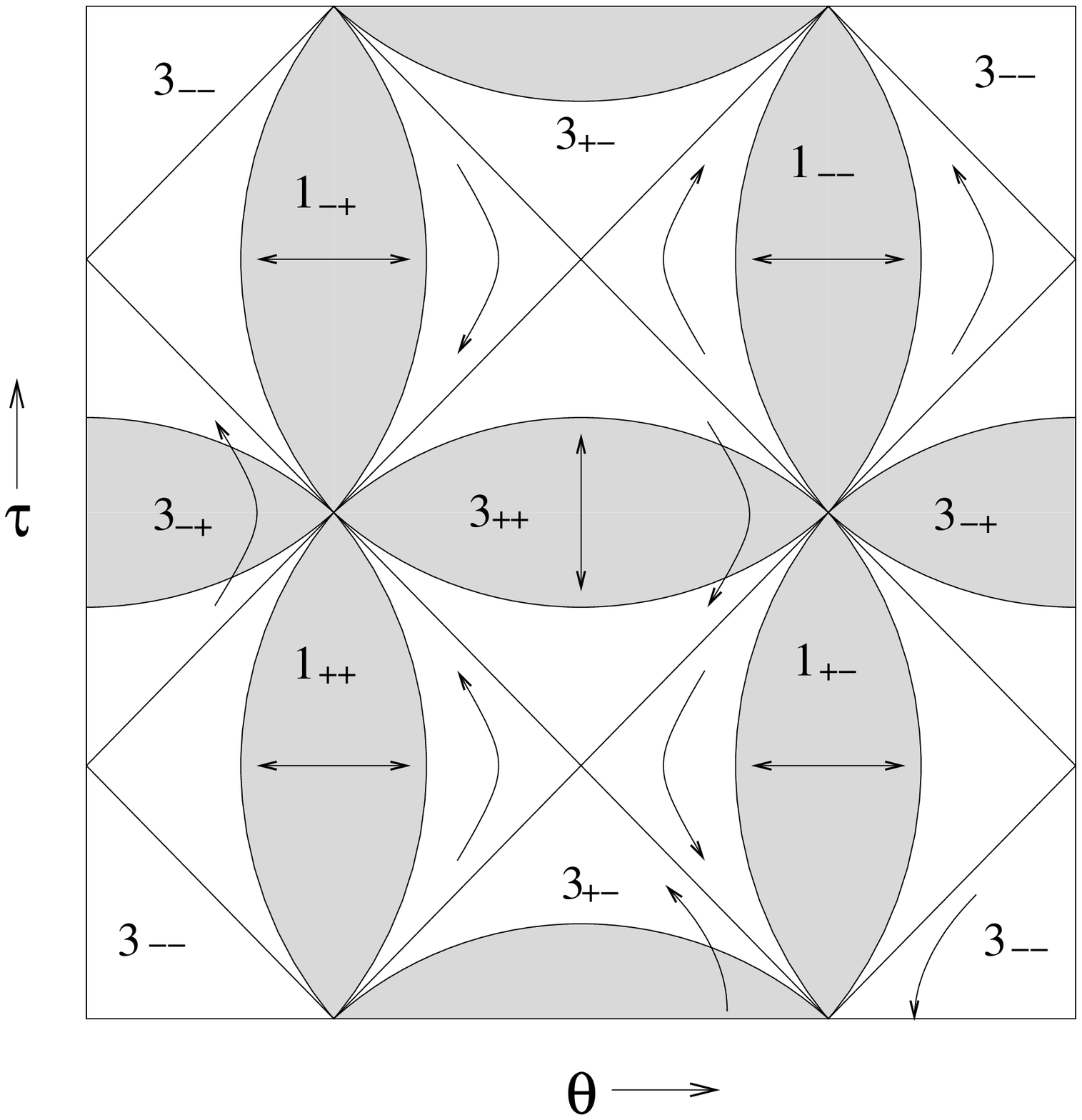}{6.0truein}
The BTZ identifications on the boundary are inherited
from \as:
\eqn\cd{\eqalign{{\rm Regions ~1,2:}\quad &
(t,\phi) ~ \cong ~ (t,\phi+2\pi) \cr
{\rm Region ~3:}\quad & (t,\phi) ~ \cong ~ (t+2\pi,\phi).
}}

For holography, it is important to identify the topology of the boundary.
From Figure 5 it appears that the boundaries of regions 1 and 3 touch
one another, but this is misleading.   From the definitions \ak\ it is
clear that it is impossible to go from region 1 to region 3, or vice
versa, without passing through region 2.  On the other hand, region 2
does not extend out to the boundary, since $r$ is bounded as $r_- \leq r
\leq r_+$.    Therefore, the BTZ boundary is disconnected.  As we will discuss
in more detail later, according to
the AdS/CFT correspondence the bulk theory is then dual to a
CFT living on the full disconnected boundary.  The main question which
needs to be addressed  is
how to relate bulk correlation functions to correlation functions of the
CFT living on this disconnected space.  We will answer this momentarily,
but  the point to be
emphasized now is that since the BTZ and AdS$_3$ spacetimes have  a
different boundary structure they correspond to distinct boundary theories;
the BTZ solution should {\it not} be thought of as a particular state in the
CFT corresponding to AdS$_3$. This is in contrast to a collapse geometry
in which  a black hole forms in the bulk; in that case the boundary is
a single connected cylinder, and we expect to be able to describe the
black hole by a pure state in the corresponding CFT.

\subsec{Euclidean black hole}

Computing string theory correlation functions directly in the Lorentzian
signature BTZ spacetime is challenging  since the worldsheet action is
unbounded from below.  In such situations one proceeds by analytically
continuing to Euclidean signature, computing correlation functions there,
and then continuing back.  This strategy was employed for pure AdS$_3$
in \MOIII\ and we wish to do the same for BTZ.

To continue we take $t \rightarrow i\tau$, or equivalently
\eqn\pa{u_+ ~ \rightarrow~ u = \phi+i\tau, \quad
u_- ~\rightarrow ~ \bar{u} = \phi -i\tau,}
where $\bar{u}$ denotes complex conjugate.   The metric \at\ then becomes
complex.  To obtain a real metric we take $r_-$ pure imaginary,
or equivalently
\eqn\pb{T_+ ~\rightarrow~ T, \quad T_- ~ \rightarrow \bar{T}.}
The metric becomes
\eqn\pc{\eqalign{ ds^2 & = { (r^2 -r_+^2)(r^2-r_-^2) \over r^2}d\tau^2
+{r^2 \over (r^2 -r_+^2)(r^2 -r_-^2)}dr^2
+r^2(d\phi - {r_+ (ir_-) \over r^2} d\tau)^2.}}
The radial coordinate  has the range $r_+ \leq r \leq \infty$.
The identifications are now
\eqn\pd{ (u,r) \cong (u+2\pi,r) \cong (u+i\beta ,r),}
where the complex Euclidean inverse temperature is $\beta = {1 \over T}$.
The boundary of the Euclidean black hole is therefore a torus with modular
parameter $i\beta /(2\pi)$.

\newsec{Supergravity correlation functions }

We now turn to the computation of correlation functions in the extended
BTZ geometry.  As compared to the usual AdS/CFT setup the novelty here
is the disconnected boundary and the presence of horizons and closed timelike
curves.   We want to establish the rules for computing correlation functions
in the bulk and boundary, as well as the relation between them.  As usual,
it is simplest to start by restricting attention to the low energy field
theory in the bulk; the full string theory is considered in  section 5.
In this section we always work in Lorentzian signature.

The simplest correlators are one-point functions,
in particular the expectation value of the boundary energy momentum tensor.
On each boundary this is given by \BalasubramanianRE\
\eqn\da{ T_{\mu\nu} = {1 \over 8 \pi G}\left( \Theta_{\mu\nu}
-\Theta \gamma_{\mu\nu} - {\gamma}_{\mu\nu}\right)}
where $\gamma_{\mu\nu}$ is the boundary metric and $\Theta_{\mu\nu}$ is
its extrinsic curvature.  Since the metric takes the form \at\ in all
regions,  the energy momentum tensor is the same
 on all boundaries:
\eqn\db{\eqalign{T_{tt} = T_{\phi\phi} & = {M \over 2\pi} \cr
T_{t\phi} = T_{\phi t}& = {J \over 2\pi},}}
with the mass and angular momentum given in \ag.  In region 1 $\phi$
has $2\pi$ periodicity, so integrating $T_{\mu\nu}$ over $\phi$ gives
the total energy and angular momentum $M$ and $J$.  On the other hand, in
region 3 $\phi$ is noncompact, so the constant $T_{\mu\nu}$ leads to
an infinite total energy and angular momentum.

Now consider two-point functions corresponding to minimally coupled bulk
scalars of mass $m$.   As usual, the two-point functions
  follow directly from the bulk-boundary propagator
\refs{\GubserBC,\WittenQJ}.    It is convenient
to start with the expression for the AdS$_3$ bulk-boundary propagator
written in Poincar\'{e} coordinates with line element,
\eqn\dc{ds^2 = { dy^2 + dw_+ dw_- \over y^2}.}
The bulk-boundary propagator behaves near the $y=0$ boundary as
$y^{2h_-} \delta^{(2)}(\Delta w_+,\Delta w_-)$,
 with $\Delta w_{\pm} = w_\pm - w_\pm'$, and corresponds to a
boundary operator of mass dimension $2h_+$, where
\eqn\dd{ h_{\pm} = {1 \over 2}(1 \pm \sqrt{1+ m^2}).}
The bulk-boundary propagator is then
\eqn\de{K_{{\rm AdS}_3}(y, w_+,w_-; w_+',w_-') = c\left(
{y \over y^2 + \Delta w_+ \Delta w_-}\right)^{2 h_+}.}
This can be rewritten in BTZ coordinates for region  $1_{++}$ using
the transformation:
\eqn\df{\eqalign{ w_\pm & =  \sqrt{{r^2 - r_+^2 \over r^2 -r_-^2}}
e^{2\pi T_\pm u_\pm} \cr
y & = \sqrt{{r_+^2 - r_-^2 \over r^2 -r_-^2}}e^{\pi(T_+u_++T_-u_-)}.}}
The resulting expression has  boundary behavior
$e^{-2\pi h_+(T_+ u_+' +T_- u_-')}r^{-2h_-}\delta^{(2)}(\Delta u_+,
\Delta u_-)$,
so we should further multiply by $e^{2\pi h_+(T_+ u_+' +T_- u_-')}$ to
get the correct BTZ propagator.   Finally, we should impose periodicity
under the BTZ identifications by including a sum over images. We first
assume that both arguments of the propagator are in region $1_{++}$, and
denote this by $K_{{\rm BTZ}}^{(1_{++} 1_{++})}$.   To
evaluate two point functions we only need the result for large $r$, which
is
\eqn\dg{\eqalign{& K_{\rm BTZ}^{(1_{++}1_{++})}(r,u_+,u_-;u_+',u_-') \cr
& \quad \quad= c'\sum_{n=
  -\infty}^{\infty}
{ \left( {r_+^2 -r_-^2 \over r^2}\right)^{h_+} e^{-2\pi
h_+[T_+\Delta u_+ + T_-\Delta u_- +(T_++ T_-)2\pi n]} \over
\left\{ {r_+^2 - r_-^2 \over r^2}+(1- e^{-2\pi T_+(\Delta u_+
+2\pi n)}) (1- e^{-2\pi T_-(\Delta u_- +2\pi n)})\right\}^{2
h_+}}. }}
Following the standard procedure gives the two-point function
\KeskiVakkuriNW\
\eqn\dh{\eqalign{\langle  & {\cal O}_{1_{++}}(u_+,u_-)
{\cal O}_{1_{++}}(u_+', u_-')\rangle\cr & \quad\quad ~\sim
\sum_{n=-\infty}^{\infty}
\left[\sinh \pi T_+( \Delta u_+ + 2\pi n)\right]^{-2h_+}
\left[\sinh \pi T_- (\Delta u_-+  2\pi n)\right]^{-2h_+}.}}

Before considering the two-point functions for operators inserted on other
boundaries, we should note that the bulk-boundary propagator \dg\ is not
the only possible choice.  As always in Lorentzian versions of the
AdS/CFT correspondence \BalasubramanianSN,
it is possible to add a solution of the bulk
wave equation with boundary behavior $r^{-2h_+}$.  \dg\ is a natural
propagator to take as it is the one which arises upon analytic continuation
from Euclidean signature.   This corresponds to evaluating expectation
values in a particular state of the dual CFT; other choices for the
propagator correspond to considering other states.

To find two-point functions on other boundaries we can use the analytic
continuation given in \ava\ and \ax.  In particular, fixing $u_\pm'$ to
be on a fixed boundary, we can use \ava\ to compute the propagator on the
full extended BTZ spacetime and then read off the resulting correlation
functions.   As above, using analytic continuation
implicitly commits one to considering a particular state of the  CFT
defined on the full disconnected BTZ boundary. We will discuss more
below whether the analytic continuation is justified.

There are eight
possible choices for inserting each of the two operators,  $1_{\pm\pm}$
and $3_{\pm\pm}$  (actually there are an infinite number of copies of
each, but given the overall periodicity these do not need to be discussed
separately).  Since only $\Delta u_\pm$ appears in \dh\ it is clear that
the same two point function is obtained whenever both operators are
inserted on the same boundary.  Now consider operators on distinct
boundaries.  Without loss of generality we can take $u_\pm '$ on boundary
$1_{++}$.   Continuing $u_\pm$ to the other boundaries then gives
\eqn\di{\eqalign{ \langle  & {\cal O}_{{\eta_1\eta_2}}(u_+,u_-)
{\cal O}_{1_{++}}(u_+', u_-')\rangle = \cr
& 1_{++}: ~ \sum_{n=-\infty}^{\infty}
\left[\sinh \pi T_+(\Delta u_+ +2\pi n)\right]^{-2h_+}
\left[\sinh \pi T_-(\Delta u_- + 2\pi n)\right]^{-2h_+}\cr
& 1_{+-}: ~ \sum_{n=-\infty}^{\infty}
\left[\cosh \pi T_+(\Delta u_+ +2\pi n)\right]^{-2h_+}
\left[\cosh \pi T_-(\Delta u_- + 2\pi n)\right]^{-2h_+}\cr
& 1_{-+}: ~ \sum_{n=-\infty}^{\infty}
\left[\cosh \pi T_+(\Delta u_+ + 2\pi n)\right]^{-2h_+}
\left[\cosh \pi T_-(\Delta u_- + 2\pi n)\right]^{-2h_+} \cr
& 1_{--}:~  \sum_{n=-\infty}^{\infty}
\left[\sinh \pi T_+(\Delta u_+ + 2\pi n)\right]^{-2h_+}
\left[\sinh \pi T_-(\Delta u_- + 2\pi n)\right]^{-2h_+} }}
and similary continuing to boundaries of the regions 3:
\eqn\dip{\eqalign{ \langle  & {\cal O}_{{\eta_1\eta_2}}(u_+,u_-)
{\cal O}_{1_{++}}(u_+', u_-')\rangle = \cr
& 3_{++}: ~ \sum_{n=-\infty}^{\infty}
\left[\cosh \pi T_+(\Delta u_+ + 2\pi n)\right]^{-2h_+}
\left[\sinh \pi T_-(u_- +u_-'- 2\pi n)\right]^{-2h_+} \cr
& 3_{+-}: ~ \sum_{n=-\infty}^{\infty}
\left[\sinh \pi T_+(\Delta u_+ +2\pi n)\right]^{-2h_+}
\left[\cosh \pi T_-(u_-+u_-' - 2\pi n)\right]^{-2h_+}\cr
& 3_{-+}: ~ \sum_{n=-\infty}^{\infty}
\left[\sinh \pi T_+(\Delta u_+ + 2\pi n)\right]^{-2h_+}
\left[\cosh \pi T_-(u_- +u_-' - 2\pi n)\right]^{-2h_+} \cr
& 3_{--}:~  \sum_{n=-\infty}^{\infty}
\left[\cosh \pi T_+(\Delta u_+ + 2\pi n)\right]^{-2h_+}
\left[\sinh \pi T_-(u_- +u_-' - 2\pi n)\right]^{-2h_+}.
 }}
These two-point functions are periodic under
\eqn\dj{( t, \phi) ~ \rightarrow ~
( t + i \beta_H, \phi + i \Omega \beta_H),}
where $\beta_H$ is the inverse Hawking temperature and $\Omega$ is
the angular velocity of the outer horizon,
\eqn\dk{\eqalign{\beta_H & = {1 \over T_H} = {T_+ + T_- \over 2 T_+ T_-} \cr
\Omega & = -{T_+ -T_- \over T_+ + T_-}.}}

\subsec{Boundary description of correlators}

In the boundary description a CFT is defined on each component of the
boundary.  We should be able to relate the above bulk correlators to
correlators in this collection of CFTs.   Since the CFTs are defined on
distinct surfaces, they will only communcate with each other via
correlations in their initial conditions and by
boundary conditions. These can be deduced from the analytic
continuations taking us from one region to another.  For operators inserted
in regions $1_{++}$ and $1_{+-}$ this was discussed in detail in
\MaldacenaKR.

Consider inserting an operator ${\cal O}$ in a given region, say $1_{++}$.
We then perform the CFT path integral over this region with boundary
conditions at $t = \pm \infty$ labelled by wavefunctionals
$\Psi_{1_{++}}(t = \pm \infty)$, yielding
\eqn\ha{ \langle \Psi'_{1_{++}}(t=\infty)| {\cal O}_{1_{++}} |
\Psi_{1_{++}}(t=-\infty) \rangle.}
We can do the same on another boundary, say $1_{+-}$.   Since the
analytic continuation \ava\ is $(t,\phi) \rightarrow (t-i\beta_H/2,
\phi - i \Omega \beta_H/2)$
the states in the two regions are related by
\eqn\hb{ |\Psi_{1_{+-}}(t=\pm\infty) \rangle ~ = ~
\langle \Psi_{1_{++}}(t=\pm\infty)| e^{-\beta_H(H- \Omega J)/2},
}
where the change from ket to bra occurs because time runs in opposite
directions in the two regions.  We will suppress the normalization factors
in state vectors.
 Therefore, the path integral in the two
regions summed over
all boundary conditions gives (up to normalization):
\eqn\hc{\sum_{E,E',J,J'}e^{-\beta_H(E+E'-\Omega J-\Omega J')/2}
\langle E,J|{\cal O}_{1_{+-}}|
E',J'\rangle
\langle E',J'|{\cal O}_{1_{++}}|
E,J\rangle .}
This result can be interpreted as follows.
First rewrite  states and operators in $1_{+-}$ in terms of their time
reversed
versions, so that time runs in the same direction in both regions.
Recall that time reversal acts as $\langle T \chi | T\psi \rangle
= \langle \psi | \chi\rangle$.
 Then consider the tensor product
of the two Hilbert spaces, and the particular correlated state
\eqn\hd{ |\Psi \rangle ~=~\sum_{E,J} e^{-\beta_H(E-\Omega J)/2}
|E,J\rangle_{1_{+-}}
|E,J\rangle_{1_{++}}.}
We find that \hc\ is equivalent to the expectation value in this state:
\eqn\he{ \langle \Psi|{\cal O}_{1_{+-}}{\cal O}_{1_{++}} |\Psi\rangle .}
If we do not insert any operator in $1_{+-}$ then we recover
a thermal expectation value for ${\cal O}_{1_{++}}$:
\eqn\hf{ \langle {\cal O}_{1_{++}} \rangle_{\beta_H,\Omega} ~=~
\sum_{E,J} e^{-\beta_H(E-\Omega J)} \langle E,J |
{\cal O}_{1_{++}}|E,J \rangle.}

This structure is very natural from the bulk point of view.
Regions $1_{++}$ and $1_{+-}$ are spacelike separated and so operators
in distinct regions commute.  Furthermore, one needs to combine
spacelike hypersurfaces in both regions in order to get a complete
spacelike hypersurface for the full spacetime.  Therefore, the full
Hilbert space is described by a tensor product of the two Hilbert
spaces, and the precise correlation in \hd\ corresponds to considering
a black hole in thermal equilibrium.  All of this then carries over
to the boundary theory.

Insertions of operators in $1_{-+}$ and $1_{--}$ are interpreted similarly.
To continue from $1_{++}$ to $1_{-+}$ we take
 $(t,\phi) \rightarrow (t-i\Omega \beta_H/2,
\phi - i  \beta_H/2)$, therefore the path integral is
\eqn\hg{
\sum_{E,E',J,J'}e^{-\beta_H(\Omega E+\Omega E'- J- J')/2}
\langle E,J|{\cal O}_{1_{-+}}|
E',J'\rangle
\langle E',J'|{\cal O}_{1_{++}}|
E,J\rangle ,}

which corresponds to an expectation value in the correlated state
\eqn\hh{ |\Psi \rangle ~=~\sum_{E,J} e^{-\beta_H(\Omega E- J)/2}
|E,J\rangle_{1_{-+}}
|E,J\rangle_{1_{++}}.}
Region $1_{--}$ corresponds to the state (in this case no time reversal
is required)
\eqn\hi{ |\Psi \rangle ~=~\sum_{E,J} e^{-(1+\Omega)\beta_H(E- J)/2}
|E,J\rangle_{1_{--}}
|E,J\rangle_{1_{++}}.}

Actually, there is a fundamental difference in the three cases we have
considered in that region $1_{+-}$ is spacelike separated in the
bulk from $1_{++}$, while  $1_{-+}$ and $1_{--}$ lie to the future
of $1_{++}$.   Since in the last two cases we are computing
correlation functions for operators inserted in causally connected
regions, it may seem unnatural to be introducing a tensor product
Hilbert space.   In the latter cases, the tensor product Hilbert space
does not correspond to the space of physical states of the theory,
but is better thought
of as a device for computing correlation functions.   According to the
AdS/CFT correspondence we should be able to relate all bulk observeables
to quantities in the boundary theory, and to do this it turns out to be
useful to employ the product Hilbert space description.  On the other
hand, from the bulk spacetime geometry it is clear that it {\it is}
sensible to combine regions $1_{-+}$ and $1_{--}$ into a tensor
product, since one thereby obtains a family of spacelike hypersurfaces.
 These hypersurfaces lie to the future of those corresponding to
region $1_{++}$ and $1_{+-}$, and so the respective
states are related by Hamiltonian
evolution.

Now consider inserting an operator in region 3.  The CFT in
each boundary component of region 3 lives on a cylinder with the
compact direction being timelike, as compared to region 1 where the
compact direction is spacelike.  We will only consider region $3_{++}$
to avoid undue repetition. The continuation from region $1_{++}$ is
$(t,\phi) \rightarrow (\phi -  i(1+\Omega)\beta_H/2,t-  i(1+\Omega)\beta_H/2)$.
The continuation maps the wavefunctions defined at early and late times
in $1_{++}$ to wavefunctions defined at spacelike infinity in $3_{++}$.
Therefore, $E$ and $J$ eigenvalues are interchanged, which is expected
since $E$ should have an integer spectrum in region $3_{++}$ due to the
timelike identification there.  Following the same logic as before, we
then find that correlators can be reproduced by taking expectation
values in the state
\eqn\hj{ |\Psi \rangle ~=~\sum_{E,J} e^{-(1+\Omega)\beta_H(E- J)/2}
|J,E\rangle_{3_{++}}
|E,J\rangle_{1_{++}}.}

We now ask whether the expectation value of operators in the
states we have defined yield the two-point functions in \di\ \dip.
In general, making the comparison would require computing
correlations functions in the strongly coupled CFT, but in a
certain limit they can be found by a combination of conformal
mappings and the method of images.   In particular, the method of
images is a useful way of imposing the correct $\phi$ periodicity,
but is only justified if correlation functions satisfy free field
equations so that solutions can be superimposed.   This occurs
when the string coupling in the bulk is taken to vanish, so in
this limit we can check that the bulk and boundary correlators
agree.

We proceed by applying an appropriate conformal transformation to
the two point function on the infinite w-plane
\eqn\hk{\langle {\cal O}(w_+,w_-) {\cal O}(w_+',w_-')\rangle
~ \sim ~ {1 \over (\Delta w_+ \Delta w_-)^{2 h_+}}.}
For instance, consider a thermal expectation value corresponding to
inserting both operators on the same boundary component.  According to
\hf\ we are to evaluate the two-point function on a torus with
identifications $(u_+,u_-)\cong (u_++2\pi, u_-+ 2\pi) \cong
(u_+-i/T_+, u_--i/T_-)$.   By defining $w_\pm$ as
\eqn\hl{w_\pm = e^{2\pi T_\pm u_\pm}}
and including a sum over images under
$w_\pm \rightarrow e^{(2\pi)^2 n T_\pm}w_\pm $
we account for the correct periodicities.  Transforming the $w$-plane
two-point function \hk\ (with the sum over images) to the $u$-frame we
recover the result in \di.  This is as it should be, since \hl\ is the
asymptotic relation between Poincar\'{e} and BTZ coordinates; see \df.
 The same story holds for the other two-point
functions except that we use a different conformal map for the two
operators. It would be interesting to repeat this analysis in the
case of nonvanishing string coupling.

\subsec{Discussion}

We have given rules for relating two-point functions in the bulk to
two-point function in the CFT defined on the disconnected BTZ boundary.
Analytic continuation allowed us to extend the bulk-boundary propagator
from one region to the full geometry, and the two-point functions then
follow in the usual fashion.   The same procedure on the boundary side
corresponds to considering various tensor product states depending on
which boundaries operators are inserted.   This procedure could clearly
be extended to higher point correlators.

An important question is whether the analytic continuation is physically
sensible.   First, as we have stressed, this procedure implicitly
chooses a particular state of the system  (the Hartle-Hawking state)
which may or may not be physically realizable.  In particular, it is
well known that the one-loop expectation value of the energy-momentum tensor
of a free scalar field in this state suffers a divergence at the inner
horizon \LifschytzEB; this is a possible mechanism for excising the
regions with closed timelike curves.  So one can argue that any
classical calculations sensitive to the geometry at the inner horizon are
unreliable. Related issues have been discussed recently in the context  of
time dependent orbifolds of flat spacetime
\refs{\BalasubramanianRY,\CornalbaFI,
\NekrasovKF,\NekrasovKF,\SimonMA,\LiuFT,\LawrenceAJ,\FabingerKR,\HorowitzMW}.

The physics at the inner horizon is unfortunately somewhat inaccesible
with current string theory technology.  The main problem is that
direct calculations are only feasible in the Euclidean signature target
space, but the inner horizon is then absent.   A priori, there are no
obvious sources of divergences in one-loop Euclidean signature
calculations  that would continue over to divergences at the inner horizon.
Indeed, note that Euclidean  BTZ is equivalent to Euclidean
``thermal AdS$_3$'' \MaldacenaBW,
and the Lorentzian version of the latter is not
expected to receive large quantum corrections.  Obviously, these
issues need to be much better understood, but since any more complete
approach should eventually  be compared with results based on the naive
classical geometry it seems useful to develop those first, as we are
 doing here.

Another issue concerns the geometry of the boundary on which the dual
CFT lives.  In BTZ coordinates it is manifest that the boundary is
conformal to a disconnected sum of cylinders, with spacelike and
timelike circle directions in regions 1 and 3.  The rules of
AdS/CFT tell us that the CFT can be taken to live on this geometry.
On the other hand since BTZ is an orbifold of \ads, and the latter has
a connected cylindrical boundary, it was proposed in \MartinecXQ\ that the
CFT should live on this cylinder  with twist operators inserted to
account for the identifications.  Note though that any connected
boundary necessarily passes through region 2 behind the horizon,
since it is not possible to go from region 1 to 3 without doing so.
We do not wish to consider boundaries traversing the horizon, and
so we prefer to work on the disconnected boundary at large radial
BTZ coordinate.

\newsec{SL(2,R) and SL(2,C)/SU(2) WZW models}

We now turn to string theory.  String theory on BTZ is described
by an orbifold of the SL(2,R) WZW model.  The spectrum of the
SL(2,R) model  was worked out in \refs{\MOI,\MOS}, and this was
extended to the BTZ orbifold in   \HemmingWE\ extending the work of
\NatsuumeIJ.
Correlation functions of the SL(2,R)
model were obtained in \MOIII\ by analytic continuation from the
Euclidean signature SL(2,C)/SU(2) model, and we would now like the
analogous story for BTZ.    We begin by reviewing the salient aspects
of the SL(2,R) and SL(2,C)/SU(2) WZW models.

\subsec{Spectrum}

The SL(2,R) WZW model is based on a $\widehat{SL}_k(2,R)_L
\times \widehat{SL}_k(2,R)_R$ current algebra,
Its Hilbert space therefore
  can be decomposed into  various irreducible
representations $\Dcal^w_j,\Ccal^w_{j,\alpha}$ of the
current algebra,
\eqn\eka{{\cal H}_{AdS_3} = \bigoplus^{\infty}_{w=-\infty} \sa \left(
 \int^{k-1 \over 2}_{\half} dj~\Dcal^w_j \otimes \Dcal^w_j
 \right) \oplus \left( \int_{\half +iR} d j \int^1_0
 d\alpha~\Ccal^w_{j,\alpha}\otimes \Ccal^w_{j,\alpha} \right)\sk}
where $\Dcal^w_j$ are the representations generated by
spectral flow from the discrete representations $D^0_j$, and
$\Ccal^w_{j,\alpha}$ are the representations generated by spectral
flow from the continuous representations $\Ccal^0_{j,\alpha}$,
with an integer spectral flow parameter $w$. States in the former
representations correspond to short strings in \ads, while states
in the latter representations correspond to long strings in \ads.

The SL(2,C)/SU(2) model has an $\widehat{SL}_k(2,C)$  current algebra,
and its Hilbert space has the structure \TeschnerFT
\eqn\ekd{{\cal H}_{H_3} =
\int_{s>0} ds~s^2~{\cal D}_{\half +is}
}
where ${\cal D}_j$
are the principal series representations
of the $\widehat{SL}_k(2,C)$  current algebra.

The connection between the SL(2,R) and SL(2,C)/SU(2) models is the following.  In
Poincar\'{e} coordinates with \ads\ metric $ds^2 = d\phi^2 + e^{2\phi}
d\gamma_+ d\gamma_-$
 the worldsheet action of the SL(2,R) model is
\eqn\ekca{ {k \over \pi} \int\! d^2z\, \left(\pat \phi \patb \phi
+ e^{2\phi} \partial \gamma_-  \bar{\partial}\gamma_+\right).}
The analytic continuation to Euclidean signature \h3\ corresponds
to $\gamma_+ \rightarrow \gamma, ~ \gamma_- \rightarrow \bar{\gamma}$,
yielding the action
\eqn\eke{
 S = {k \over \pi} \int\! d^2z \,\left(\pat \phi \patb \phi +
e^{2\phi} \bar{\partial}\gamma \pat \bar{\gamma}\right).}
On the other hand, one can
regard the coset SL(2,C)/SU(2) as the space of matrices $g$
parametrized by
\eqn\ekf{
   g = \pmatrix{e^{-\phi}+\gamma \gammab e^\phi &
    e^\phi \gamma \cr e^\phi \gammab & e^\phi } \
    .}
This follows from the fact that $g= h h^\dagger$ with $h \in$ SL(2,C).
Then, substituting $g$ into the standard form of the WZW
action yields the action in \eke.

The analytic continuation of the target space time coordinate creates
subtleties  when making connections with the two
models. The first issue is associated with the
normalizability of the associated string states.
In the worldsheet theory,
states in the Schrodinger picture are wavefunctionals $\Psi
[x^\mu (\sigma )]$
where $x^\mu$ are the target space coordinates including the target
space time coordinate $t$. Worldsheet normalizability is defined
by integrating $|\Psi|^2$ over all $x^\mu (\sigma )$, so the
target space time dependence enters into the normalizability
condition.

From the worldsheet point of view all states in the Hilbert space
of either theory are normalizable; note that  we should allow for delta
function normalizability due to the noncompact target space.
We can choose a basis of states with time
dependence $e^{-i \omega t}$, where $t$ is the zero mode of the Lorentzian
or Euclidean Poincar\'{e} time coordinate.   Normalizability then requires
boundary behavior $|\Phi| \sim e^{-p\phi},~ {p \geq 1}$.
Spin $j$ primaries obey
the wave equation for a minimally coupled scalar field
\eqn\zaa{\left( \Delta - m^2 \right) \Phi_j =0, \quad m^2 = 4j(j-1),}
where $\Delta$ is the Laplacian on SL(2,R) or SL(2,C)/SU(2).
This equation has solutions with boundary behavior
\eqn\za{ \Phi_j ~ \sim ~ \left\{\eqalign{ & e^{-2h_\pm \phi }
\quad {\rm Lorentzian} \cr
& e^{-2h_-\phi} \quad   {\rm Euclidean}} \quad\quad  h_{\pm}= {1 \over 2}
(1 \pm \sqrt{1 +m^2})\right..}
In the Lorentzian case we can take any real $m^2$ and the $h_+$ branch,
yielding normalizable primaries for any real $j$ (but given the form
of $m^2$ we can restrict to $j>{1 \over 2}$) or $j = {1 \over 2}+ is$.
In the Euclidean case we need
$m^2 \leq -1$ or $j = {1 \over 2} + is$.   In the Lorentzian case with
real $j$ there is also the upper bound $j=(k-1)/2$ which can be understood
as the transition from short strings to long strings \MOI.
This accounts for the range of allowed $j$'s in \eka\ and \ekd.

What can be computed directly  are correlation functions of the
normalizable $j = {1 \over 2} + is$ vertex operators in the
SL(2,C)/SU(2) model.  The trouble is that in the AdS/CFT correspondence
correlation functions in the boundary theory are related to correlation
functions of non-normalizable vertex operators in the worldsheet CFT.
These correspond to taking $j$ real in the SL(2,C)/SU(2) model, or
$j$ real and $h=h_-$ in the SL(2,R) model.  Such operators   transform
in nonunitary representations of SL(2,C) and SL(2,R).  The strategy
pursued in \MOIII\ was to start from the $j = {1 \over 2} + is$
correlators of the SL(2,C)/SU(2) model, and then analytically continue
in $j$ and the target space time coordinate  in order to obtain
correlation functions of non-normalizable vertex operators in the
SL(2,R) model.  A sensible picture was obtained, with various singularities
in the $j$-plane given physical interpretations.  We follow a similar
strategy, the difference being that we will focus on vertex operators
consistent with the BTZ identifications, and we will analytically continue
to the full extended  Lorentzian BTZ geometry.

\subsec{Vertex operators in the SL(2,C)/SU(2) model}

The primary
operators of the  model fall into the unitary irreducible
representations of  SL(2,C),
labelled by $j=\half +is$ where $s$ is a real number. In the
semi-classical large k limit normal ordering issues can be neglected,
and the primaries correspond to a complete set of normalizable solutions
of \za \TeschnerFT,
\eqn\ekj{
  V_j(z,\zbar ;x,\xbar) = {1-2j \over \pi} \left( (\gamma -x)(\gammab -\xbar )
  e^\phi + e^{-\phi} \right)^{-2j} \ . }
The primaries are parameterized by the complex variable $x$; \ekj\ is
just the usual bulk-bounday propagator in \h3\ with $x$ the coordinate
on the boundary.
More generally, the vertex operators must  satisfy
the correct operator product expansions with the SL(2,C) currents,
\eqn\ekg{\eqalign{
  J^a(z)V_j(z',\zbar';x,\xbar ) &\sim  {1 \over z-z'}{\cal D}^a_j
  V_j(z',\zbar';x,\xbar ) \cr
  \Jbar^a(\zbar ) V_j(z',\zbar';x,\xbar ) &\sim  {1 \over \zbar-\zbar'}
  \bar{{\cal D}}^a_j
  V_j(z',\zbar';x,\xbar )}}
where ${\cal D}^a_j,\bar{\cal D}^a_j$ are the representation
of the SL(2,C) algebra generators, acting in the space of
functions on $C$,
\eqn\ekh{
 {\cal D}^-_j = x^2{\pat \over \pat x} +2jx,\ {\cal D}^+_j =
 {\pat \over \pat x},\ {\cal D}^3_j = x{\pat \over \pat x}+j \ . }
The operators are primaries with respect to the
Virasoro algebra, with conformal weights $\Delta_j$ given by the
Casimir $-j(j-1)$ of the representation,
\eqn\eki{
   \Delta_j = -{j(j-1) \over k-2} \ .}

The two-point functions for the vertex operators \ekj\
were computed in \refs{\TeschnerFT} and found to have the form
\eqn\ekna{
 \bra V_j(z_1,\zbar_1 ;x_1,\xbar_1)V_{j'}(z_2,\zbar_2
 ;x_2,\xbar_2)\ket
 = {1\over |z_{12}|^{4\Delta_j}} \left[
 \delta^2(x_{12})\delta (j+j'-1) +
 {B(j)\over |x_{12}|^{4j}}\delta (j-j') \right]}
where the coefficent $B(j)$ is
\eqn\eko{
  B(j) = {k-2 \over \pi} {\nu^{1-2j}\over \gamma
  ({2j-1 \over k-2})}}
and
\eqn\ekp{
  \gamma (x) \equiv {\Gamma (x)\over \Gamma (1-x)}, \quad
\nu \equiv \pi {\Gamma (1-{1\over k-2})\over \Gamma
  (1+{1\over k-2})}.
}
Correlation functions on the worldsheet correspond to
correlation functions on the boundary CFT by the relation \deBoerPP
\eqn\ekr{
 {\bra \prod_i \int d^2z_i V_{j_i}(z_i,\zbar_i ;x_i,\xbar_i)
 \ket_{ws}  \over {\rm Vol(SL(2,C))}}
 = \bra \prod_i V_{j_i}(x_i,\xbar_i )\ket_{BCFT} \ .}
For three-point or higher correlation functions one can cancel three
of the $d^2z$ integrals against Vol(SL(2,C)) to get a finite result
on the left hand side.
But for the two-point function one is still left in the denominator
with the volume of the SL(2,C) subgroup leaving two points
fixed (the dilatation group).
The volume is infinite, but it can cancel the
divergence coming from the delta function $\delta(j-j')$ \KutasovXU.
In the process
a $j$ dependent factor can appear, but this can be fixed by relating
the result to a three-point function using a Ward identity \MOIII.
One thereby obtains
\eqn\eks{
  \bra V_j(x_1,\xbar_1) V_j(x_2,\xbar_2 )\ket_{{\rm BCFT}} =
  {(2j-1)B(j)\over |x_{12}|^{4j}} \ .}

It is also useful to employ a momentum space basis for the primaries,
using the transformation
\eqn\ekm{
  V_j(z,\zbar ;x,\xbar ) = \sum_{m,\mbar}
  V_{j;m,\mbar}(z,\zbar )~x^{m-j}\xbar^{\mbar -j}}
and the inverse  transformation
\eqn\ekn{
  V_{j;m,\mbar}(z,\zbar ) = {1 \over 4 \pi^2}
  \int\! d^2x\,|x|^{-2}x^{j-m}\xbar^{j-\mbar}
  V_{j}(z,\zbar ;x,\xbar ) \ .}
One finds
\eqn\ekt{
 \bra V_{j,m,j,\mbar} V_{j,m',j,\mbar'} \ket_{\rm BCFT} =
 \delta^2(m+m') {\pi \Gamma (1-2j)\Gamma (j+m)\Gamma
 (j-\mbar)\over
 \Gamma (2j)\Gamma (1-j+m)\Gamma (1-j-\mbar)}~(2j-1)B(j) \ .}
\ekt\ can now be intepreted as a two point function in
 Lorentzian \ads\ by analytically continuing in $j$.   The pole
structure in the $j$-plane was given a physical interpretation in
\MOIII.

As stated, \ekt\ gives the two-point function for  $w=0$ operators
without spectral flow. Let us then consider two-point functions
for spectral flowed states. Spectral flow acts on the current
algebra generators and Virasoro operators as follows:
\eqn\ekua{\eqalign{
 J^3_n & = \Jtilde^3_n +{k\over 2}w\delta_{n,0} \cr
  J^{\pm}_n & = \Jtilde^\pm_{n\mp w} \cr
 L_n & = \tilde{L}_n -w\Jtilde^3_n-{k\over 4}w^2\delta_{n,0}\ .}}
Following \MOIII, we interpret the generators with tildes on top
as those of the unflowed basis and without tildes as those of the
spectral flowed basis; the notation for the eigenvalues is $j,m,\Delta_j$ in the
unflowed basis and $J,M,\Delta_J$ in the flowed basis.

In \MOIII, several spectral flowed two-point functions were
discussed. The simplest case is when the vertex operators in the
$J,M$ basis create lowest or highest weight states $J=M$ in the
representations $d^\pm_J$. A generic state in the unflowed
representation $d^\pm_j$ can always be mapped to such states by a
suitable amount of spectral flow. Since one already knows the
2-point function \ekt\ in the unflowed basis, one can then obtain the
2-point function for the $J=M$ states simply by
replacing the labels:
\eqn\ekub{m=M-{k \over 2}w \,, \quad \mbar = \Mbar -{k\over 2}w \ . }
Similarly, in the worldsheet 2-point function, one only needs to
modify the powers of $z,\zbar$ by replacing the correct conformal
weights:
\eqn\ekuc{\Delta_j \rightarrow \Delta_J = \Delta_j -wm - {k\over 4}w^2 \,, \quad
\bar{\Delta}_J = \Delta_j -w\mbar - {k\over 4}w^2 \ .}
The worldsheet two-point function ends up being
\MOIII
\eqn\ekv{\eqalign{
 &\bra
 V_{J,M,\Jbar,\Mbar}(z_1,\zbar_1)V_{J',M',\Jbar',\Mbar'}(z_2,\zbar_2)\ket  \cr
&~~~= {\delta^2(M+M')\over z_{12}^{2\Delta_J}\zbar_{12}^{2\bar{\Delta}_J}}
\left( \delta (j+j'-1) + \delta (j-j') {\pi B(j)\over \gamma (2j)}{\Gamma
(j+m)\Gamma
 (j-\mbar)\over \Gamma (1-j+m)\Gamma (1-j-\mbar)}\right)}}
with the eigenvalues related by \ekub, \ekuc. Specific choices for
the spin $j$ then give two-point functions for either spectral flowed short or
long strings; the resulting expressions can be found in \MOIII.

\newsec{Two-point functions for strings in BTZ}

Previously we discussed  2-point functions in BTZ in the supergravity
approximation.  Now we want to do the same in the full string theory. In
other words, we will now generalize the results of \MOIII, as reviewed
in the previous section,  to BTZ  black holes.
The starting point of the discussion in section 4 was
the vertex operators \ekj\ which transformed as tensors of
conformal weight $(j,j)$ on the boundary. Semi-classically, the vertex
operators \ekj\ were identified with the bulk-boundary propagator.
As discussed in section 3, the BTZ bulk-boundary propagator
is obtained by transforming the \ads\ propagator and including a sum
over images.  Its asymptotic form with both arguments in region $1_{++}$
is written in \dg; it is extended to the other regions by analytic
continuation.
If we then replace the bulk coordinates
$r,u_\pm$ with the embedding of the string world sheet
$r(z,\zbar),u_\pm(z,\zbar)$, we can interpret the bulk-boundary
propagator as the weight $(j,j)$ string vertex operator labelled by
$u'_\pm$,
\eqn\btzb{V_j(z,\zbar;u'_+,u'_-) = K_{\rm
BTZ}(r(z,\zbar),u_+(z,\zbar),u_-(z,\zbar);u_+',u_-')\
.}
Being an orbifold of \ads,  string theory on BTZ also has twisted sector
vertex operators.  These are included in the spectral flow operation,
to be reviewed below.

\subsec{String spectrum in BTZ}

Let us now review some facts about the spectrum of strings in
BTZ \NatsuumeIJ\ \HemmingWE. To build the Hilbert space, we again
start from the $\widehat{SL}_k(2,R)_L \times \widehat{SL}_k(2,R)_R$
current algebra.    For \ads\ it is convenient to work in the elliptic
basis, which includes translation generators for the global coordinates
$\tau$ and $\theta$. For
strings in BTZ black holes we instead want translation generators for
BTZ $t$ and $\phi$.  This is the hyperbolic basis
and corresponds to diagonalizing  the non-compact
generators $J^2_n,\Jbar^2_n$.  In the hyperbolic basis
the current algebra commutation relations
take the form
\eqn\btzc{\eqalign{
   \left[ J^2_n, J^{\pm}_m \right] &= \pm i J^{\pm}_{n+m} \cr
   \left[J^+_n, J^-_m \right] &= -2iJ^2_{n+m} -kn\delta_{n+m,0} \cr
   \left[J^2_n,J^2_m \right] &= {k\over 2} n\delta_{n+m,0}\, .
}}
Time translations and
rotations in BTZ coordinates are generated by the following
combinations of zero modes of $J^2,\Jbar^2$:
\eqn\btzd{\eqalign{
  Q_t  &= 2\pi T_+ J^2_0 -2\pi T_- \Jbar^2_0 \cr
  Q_\phi &= 2\pi T_+ J^2_0 +2\pi T_- \Jbar^2_0 }}
(up to additional constant terms for winding modes, see
\HemmingWE). The eigenvalues $J_L,J_R$ of the non-compact global
SL(2,R) generators $J^2_0,\Jbar^2_0$ have a continuous spectrum.
That, along with the form of the commutation relations in the
hyperbolic basis, make it slightly more complicated to recognize
the standard discrete and continuous unitary irreps of the global
SL(2,R)$\times$SL(2,R) algebra. For details, see \refs{\HemmingWE, \NatsuumeIJ}. After
constructing the standard representations of the global algebra,
one can proceed to
construct the representations of the current algebra. The states
in the representations have the form \NatsuumeIJ
\eqn\btze{
     K_N |J_R ,r \ket \bar{K}_N |J_L, r\ket }
where $K_N$ is a generic product of operators $K^a_{-n}$ defined
by
\eqn\btzf{
   K^2_{-n} \equiv J^2_{-n}, \quad K^+_{-n} \equiv J^+_{-n}J^-_0,
\quad K^-_{-n} \equiv J^-_{-n}J^+_0 \ . }
These satisfy the commutation rules
\eqn\btzg{
   \sa J^2_0,K^a_{-n} \sk = 0, \quad \sa L_0 , K^{\pm}_{-n} \sk =
   nK^{\pm}_{-n} \ . }
Spectral flow in the hyperbolic basis generates strings which wind
around the horizon of the BTZ black hole. The action on the group
elements is
\eqn\btzh{ g \rightarrow e^{-i w_+ x^+ \tau^2}\, g\, e^{i w_- x^-
\tau^2}.}
In particular, the BTZ coordinates transform as
\eqn\btzia{u_\pm(\sigma,\tau) ~ \rightarrow ~ u_\pm(\sigma,\tau)
 + {w_\pm \over 2\pi T_\pm}
(\sigma \pm \tau).}
After the periodic identifications which make the BTZ black hole,
the spectral flow parameters are constrained to discrete values
\eqn\btzj{
      w_{\pm} = 2\pi T_\pm n \ ,}
so as to respect the periodicity of the worldsheet. Under spectral
flow, the components of $J^2,\Jbar^2$ transform as
\eqn\btzk{\eqalign{ J^2_n &~\rightarrow~ \tilde J^2_n \equiv J^2_n +
{k\over 2} \, w_+ \delta_{n,0}\,,\quad J^\pm_n \ ~\rightarrow ~\
\tilde J^\pm_n \equiv J^\pm_{n\pm iw_+}\cr \bar J^2_n &~\rightarrow~
\tilde{\bar{J}}^2_n \equiv \bar J^2_n - {k\over 2} \, w_-
\delta_{n,0}\,,\quad  \bar J^\pm_n \ ~\rightarrow~ \
\tilde{\bar{J}}^\pm_n \equiv \bar J^\pm_{n\pm iw_-} }}
and the Virasoro generators are found to transform as
\eqn\btzl{\eqalign{ L_n &~\rightarrow~ L_n + w_+ J^2_n + {k\over 4}
w_+^2 \delta_{n,0} \cr \bar L_n &~\rightarrow~ \bar L_n - w_- \bar
J^2_n + {k\over 4} w_-^2 \delta_{n,0} \ . }}
The vertex operators for the Kac-Moody primaries, tranforming
under the unitary irreducible representations of the global
algebra, have the form
\eqn\btzm{
 V^{j,0}_{J_R,J_L} = D^j_{J_R,J_L} (g) e^{-iJ_R\hat{u} +iJ_L \hat{v}} \ ,}
where $\hat{u}=\pi T_+u_+,\ \hat{v}=-\pi T_-u_-$.
The vertex operators $V^{j,n}_{J_r,J_L}(z,\zbar)$ in the twisted
sector are
constructed with the help of twist fields $W_n(z,\zbar)$,
\eqn\btzn{
 V^{j,n}_{J_R,J_L}(z,\zbar) = V^{j,0}_{J_R,J_L}(z,\zbar )
    W_n(z,\zbar) \ . }
The Kac-Moody primaries are then the states
\eqn\btzo{
   | j,J_R,n\ket | j, J_L,n \ket = V^{j,n}_{J_R,J_L}|0\ket|0\ket \
   .}
Alternatively, the twisted sector states can be interpreted as
the spectral flowed states with integer flow parameter $n$.\foot{In 
addition, there is one subtle
part of the spectrum which was not discussed in  
\refs{\NatsuumeIJ,\HemmingWE}, the discrete set of vacua \SatohXE.
These presumably correspond to the vacua defined with respect to each
spectral flowed basis of the current algebra. We thank Y. Satoh
for bringing this to our attention.}

The target space energy spectrum for the string states is
continuous, unlike in pure \ads\ where the spectrum of short strings
 is discrete. The
target space energy is  given by the eigenvalue of the time
translation generator $Q_t$, which involves continuous eigenvalues
for the operators $J^2_0,\Jbar^2_0$. More discussion can be found
in \HemmingWE.

The preceding construction gives us the spectrum of strings located
in, say, region $1_{++}$ outside the horizon.  In what sense are the
strings localized in a given region, and how do we obtain the string
states
in the other regions?    We chose to diagonalize the zero mode
generators in the hyperbolic basis; these
act as isometry generators in a given  coordinate patch of
the BTZ spacetime.   Therefore,
the center of mass coordinates of the string are confined to a given
patch.   On the other hand,
the full wavefunction of the string spreads out into the other
regions.  This becomes clear if one thinks
of starting with a string state in \ads\ and then imposing the BTZ
identifications ---  the original state
spreads out over the whole \ads\ spacetime, and the identifications
affect only the center of mass
coordinates, thus the final state spreads out over the whole BTZ
spacetime.   Our spectrum therefore
includes strings that straddle the horizon, a picture which is
reminiscent of strings ending on a D-brane,
and which has been suggested to be responsible for black hole entropy
(for discussion, see e.g. \SusskindQC).
The other issue concerns strings
with center of mass coordinates in the different regions.   Since the
WZW model takes the same form in
all regions, it is clear that the analysis is essentially identical in
all cases.  The only difference is that
the coordinate range of $r$ is different in region 2 from regions 1
and 3, but this just affects one's choice
for a complete basis of solutions to the wave equation in a given
region.  So modulo this fact, we get the
same spectrum of strings in all regions.   Actually, in order for
interactions to be well behaved the
center of mass wavefunctions should continue smoothly from one region
to a neighboring region.
Exactly as in field theory, this can correlate positive and negative
frequency wavefunctions in
adjacent regions,  and is responsible for Hawking radiation.

\subsec{Identifying the vertex operators}

In order to relate the vertex operators proposed in \btzb\ with
the Kac-Moody primaries \btzm, we need to first transform the
former from the $(j,u'_+,u'_-)$ basis to the $(j,J_R,J_L)$ basis. Note
first that the operators \btzb\ included a sum over
images which rendered them periodic in $\phi$ --- they are of the
form
\eqn\btzp{V_j(z,\zbar;u'_+,u'_-) = \sum^\infty_{n=-\infty}
\tilde{V}_j(z,\zbar;u'_++2\pi n,u'_-+2\pi n)\ , }
where $\tilde{V}_j$ is equal to \dg\ without the sum. We can
express them as Fourier integrals (we simplify the notation and
drop the primes from the boundary coordinates)
\eqn\btzppa{V_j(z,\zbar;u_+,u_-) = \sum^\infty_{k=-\infty}
\int d\omega~V_{j,\omega,k}~e^{+i({\omega - k\over 2})u_+
-i({\omega +k\over 2})u_-}\ , }
with the inverse transfomation
\eqn\btzppb{V_{j,\omega ,k} = {1 \over 4 \pi^2}
\int du_+ \int du_-~e^{-i({\omega - k\over 2})u_+
+i({\omega +k\over 2})u_-}V_j(z,\zbar;u_+,u_-)\ .}
On the boundary, $J^2_0,\Jbar^2_0$ are represented by
\eqn\btzppc{\eqalign{D^2 &= -i{1\over 2\pi T_+}\pat_{u_+} \cr
\bar{D}^2 &= -i{1 \over 2\pi T_-}\pat_{u_-}\ . }}
The Fourier modes are their eigenfunctions with
eigenvalues
\eqn\btzppd{J_R = {\omega - k\over 4\pi T_+},\quad
 J_L = -{\omega +k\over 4\pi T_-} \ .}
The $J_R,J_L$ have continuous real eigenvalues as expected, and thus
 we can denote the vertex operators in Fourier space by $V_{j,J_R,J_L}$ as in
the previous section.

We will now move on to compute the 2-point functions for strings
in BTZ. The calculations are based on those in \TeschnerFT\ and
\MOIII, so we need to perform an analytic continuation to Euclidean BTZ
geometry.

\subsec{Euclidean section}

As in Section 3, we are going to work in the Euclidean geometry.
The Euclidean section for the BTZ black hole is obtained by
\eqn\btzpa{\eqalign{u_+ &\rightarrow u = \phi +i\tau, \quad u_-\rightarrow
\ubar = \phi -i\tau, \cr
T_+ &\rightarrow  T, \quad  T_- \rightarrow  \bar{T}.}}
The boundary of the Euclidean BTZ is a torus
\eqn\btzq{ u \sim u +2\pi, \quad u \sim u + i\beta}
where $\beta=1/T=\beta_1+i\beta_2$ is the complex inverse temperature.
Now we need to Fourier expand the vertex operators
$V_j(z,\zbar;u,\ubar)$ with mode
functions $f_{m,\mbar}$ which are periodic on the torus,
\eqn\btzr{ f_{m,\mbar}(\phi ,\tau) = e^{i(m-\mbar)\phi + {i\over
\beta_1}[(2\pi +\beta_2)m +(2\pi -\beta_2)\mbar]\tau} \ ,}
the expansion is
\eqn\btzs{ V_j(z,\zbar;u,\ubar) = \sum_{m,\mbar }
V_{j,m,\mbar}~f_{m,\mbar}(\phi ,\tau) }
and the inverse transformation is
\eqn\btzt{ V_{j,m,\mbar} = \int^{2\pi}_0 d\phi \int^{\beta_1}_0
{d\tau \over \beta_1} f_{-m,-\mbar}(\phi,\tau)~V_j(z,\zbar;u,\ubar) \ . }
The functions $f_{m,\mbar}$ are of course again eigenmodes of the
generators $J^2_0,\Jbar^2_0$. On the Euclidean boundary, $J^2_0,\Jbar^2_0$
are represented by
\eqn\btzu{\eqalign{D^2 &= -i{\beta\over 2\pi}\pat_u \cr
\bar{D}^2 &= -i{\bar{\beta}\over 2\pi}\pat_{\ubar}\ . }}
The functions $f_{m,\mbar}$ satisfy
\eqn\btzv{\eqalign{ D^2 f_{m,\mbar}(u,\ubar)&= -{i\beta\over 4\pi
\beta_1}[(2\pi +i\bar{\beta})m +(2\pi
-i\bar{\beta})\mbar]~f_{m,\mbar}(u,\ubar) \equiv
+i\alpha~f_{m,\mbar}(u,\ubar) \cr \bar{D}^2 f_{m,\mbar}(u,\ubar)&=
{i\bar{\beta}\over 4\pi \beta_1}[(2\pi -i\beta )m +(2\pi
+i\beta )\mbar ]~f_{m,\mbar}(u,\ubar) \equiv
-i\bar{\alpha}~f_{m,\mbar}(u,\ubar) \ . }}
Thus, in the Euclidean geometry the eigenvalues $J_R,J_L$ take
the values
\eqn\btzw{\eqalign{J_R &= i\alpha = -{i\beta\over 4\pi
\beta_1}[(2\pi +i\bar{\beta})m +(2\pi -i\bar{\beta})\mbar] \cr
J_L &= -i\bar \alpha  = {i\bar{\beta}\over 4\pi \beta_1}[(2\pi -i\beta
)m +(2\pi +i\beta )\mbar ] \ . }}
Note that now the eigenvalues $J_R,J_L$ take discrete
complex values. This is a
property of the Euclidean section, following from the
double periodicity of the mode functions \btzr\ on the Euclidean
toroidal boundary. In analytically  continuing back to the Lorentzian
section, the mode functions \btzr\ need to be continued to those
in \btzppa\ which are only periodic in the angle coordinate.
Therefore, in addition to continuing back to real time
coordinate, we also need to continue the parameters so that the eigenvalues
\btzw\ again take the form \btzppd.   Also, we first analytically continue
both arguments of the two-point function to the same boundary component.
Two-point functions on distinct boundary components are then obtained by
further analytic continuation.
We comment on these issues further after
we have computed the 2-point function.

\subsec{Two-point functions}

Let us now label the vertex
operators as $V_{j,i\alpha,-i\alphab}$,
and compute their two-point functions, the analogue of \ekt. It
turns out that we can use simply use the earlier results
discussed in Section 3, if we express the
inverse Fourier transformation formula \btzt\ in the coordinates
\eqn\btzx{x = e^{{2\pi\over \beta}u}, \quad
\xbar = e^{{2\pi\over \betab}\ubar} \ ,}
remembering that $V$ transforms like a tensor. It then takes a similar form
to \ekn,
\eqn\btzy{
  V_{j;m,\mbar}(z,\zbar ) = {1 \over 8 \pi^2 } \left( {4\pi^2\over \beta \betab}
  \right)^{j-1}
  \int\! d^2x\,|x|^{-2}x^{j+\alpha}\xbar^{j-\alphab}
  V_{j}(z,\zbar ;x,\xbar ) \ .}
Since in the $x,\xbar$ space the 2-point function is the same as
\eks, the result now has the same form as \ekn, but with the
replacement
\eqn\btzaa{m\mapsto -\alpha \ , \ \mbar \mapsto \alphab \ .}
So we obtain
\eqn\btzab{
 \bra V_{j,i\alpha',j,-i\alphab'} V_{j,i\alpha,j,-i\alphab} \ket_{\rm BCFT} =
 \delta^2(\alpha' +\alpha) {\pi (2j-1) B(j) \Gamma (1-2j)\Gamma (j+\alpha)\Gamma
 (j+\alphab)\over
 \Gamma (2j)\Gamma (1-j+\alpha)\Gamma (1-j+\alphab)}\ .}
Note that we only know the two-point function at a discrete set of
points $\alpha ,\alphab$ given by \btzw. This is analogous to what
happens in finite temperature quantum field theory. Usually
the thermal Green's functions are calculated in the imaginary time
formalism, and they are found only at the discrete set of Matsubara
frequencies. To understand the result \btzab\ better, let us
compare it with the two-point function previously
obtained from a supergravity BTZ/CFT calculation \DanielssonZT.
Ref. \DanielssonZT\ calculated the retarded CFT propagator in momentum
space for a CFT operator which is dual to a supergravity field in
the non-rotating BTZ black hole bulk geometry. The momentum space retarded
propagator is defined by a Fourier transformation of the
real time retarded propagator,
\eqn\btzac{
G_{ret}(\omega ,k) = \int \! dtd\phi~e^{-i\omega t+ik\phi}~G_{ret}(t,\phi)}
and the result \DanielssonZT\ is
\eqn\btzad{
 G_{ret}(\omega ,k) =
  {\Gamma (1-\nu )\Gamma (h_+-{i\over 2r_+}(\omega + k))\Gamma
 (h_+-{i \over 2r_+}(\omega -k))\over
 \Gamma (1+\nu )\Gamma (h_- -{i\over 2r_+}(\omega + k))
 \Gamma (h_- -{i\over 2r_+}(\omega - k))}\ ,}
where $h_+=j,~h_-=1-j$ and $1+\nu = 2j$.\foot{Note: eqn. (16)
in \DanielssonZT\ has a typo,
$R$ should be $R^2\over 2r_+$, we use $\ell = R$ and set $\ell
=1$.}
Comparing the
Fourier transformation formulas \btzppd\ and \btzt, we can
establish the analytic continuation of the parameters
\eqn\btzae{\alpha \leftrightarrow -{i\over 2r_+}(\omega - k), \quad
\alphab \leftrightarrow -{i\over 2r_+}(\omega + k).}
With this relation, the propagator \btzab\ has the same structure
as \btzad. More precisely, the retarded
CFT propagator \btzad\ is a Fourier transform finite temperature propagator
in real time. Hence it is known at a continuous set of frequencies
$\omega$. It is known that if we analytically  continue it
to the complex $\omega$ plane, and evaluate it at the discrete set of
Matsubara frequencies, it is equal to the finite temperature
thermal propagator calculated in the imaginary time formalism.
For a brief review, see e.g. the Appendix of \EvansFR.
This is what happens here too. From \btzae, the frequencies are
\eqn\btzaf{\omega_n = ir_+(\alpha +\alphab )=-ir_+(m+\mbar) \equiv
 -i2\pi n T_H \ ,}
which are just the expected
Matsubara frequencies.\foot{We denote $n=m+\mbar$.} Note also that
\eqn\btzae{k= -ir_+(\alpha -\alphab) = -m+\mbar = {\rm integer} .}

Consider then the pole structure of \btzab.  There are two types
of poles in the $j$-plane.  First, there are poles arising from the
$\Gamma (j-\alpha)\Gamma (j-\alphab)$ factors. These have a particle-like
interpretation as a finite temperature effect
due to the thermal density matrix of the boundary theory. Second,
there are poles arising from the factor $B(j)$. In \MOIII\ these were
interpreted as
worldsheet instantons arising from a holomorphic map of the spherical
worldsheet
onto the spherical boundary of  Euclidean \ads.
Euclidean BTZ has a toroidal boundary,
so at first sight one might
be puzzled by the absence of a holomorphic map from
a sphere
to a torus. But recalling that  the torus arises from the periodic
identifications,  we still have the worldsheet instantons but now they wrap
the boundary torus multiple times.

In section 3 we discussed BCFT correlation functions on the extended boundary. It is
straightforward to extend this discussion to string theory. We need to again
transform
from the  $J_R,J_L$ basis to the
$u_+,u_-$-basis. First, the vertex operators \btzb\ can be
analytically continued to any of the other
regions $1\eta_1\eta_2, 3\eta_1\eta_2$ by a suitable analytic continuation
of the boundary coordinates $u'_+,u'_-$. In order to compute a BCFT 2-point function
associated with a pair of such string states, it is convenient to start from the
expression \eks, use the transformation \btzx\ and remember that the operators
scale like $(j,j)$-tensors. In this way, one obtains for example the BCFT two-point
function
\eqn\btzaea{\eqalign{\bra
V^{1++}_j&(u_{+1},u_{-1}) V^{1++}_j(u_{+2},u_{-2} )\ket_{{\rm
BCFT}} \cr
 & =  \!
  \sum^{\infty}_{n=-\infty}
{(2j-1)B(j)\over \left[\sinh \pi T(\Delta u_+ +2\pi n)\right]^{2j}
\left[\sinh \pi T(\Delta u_- + 2\pi n)\right]^{2j} } }}
Continuation to other regions is the same as in \di\ \dip.

\subsec{Spectral flowed 2-point functions}

Having obtained the two-point function for unflowed operators, we
proceed as in Section 4 and compute the spectral flowed two-point
function. For convenience, we first alter some of our notations
somewhat. Let us denote the eigenvalues of $J^2_0,\Jbar^2_0$ in
the unflowed basis as $J_R,J_L$ and in a flowed basis as ${\bf J}_R,{\bf
J}_L$. We first write the worldsheet two-point function, and use
the labels $J_R,J_L$ instead of $\alpha ,\alphab$:
\eqn\spfaa{\eqalign{
 \bra &V_{j,J'_R,j,J'_L}(z_1,\zbar_1 ) V_{j,J_R,j,J_L}
 (z_2,\zbar_2 )\ket = | z_{12} |^{-4\Delta_j }~\delta^2(J'_R+J_R)~\times \cr
 & \left( \delta(j+j'-1)
 + \delta (j-j'){\pi (2j-1) B(j) \Gamma (1-2j)\Gamma (j-iJ_R)\Gamma
 (j+iJ_L)\over
 \Gamma (2j)\Gamma (1-j-iJ_R)\Gamma (1-j+iJ_L)} \right) \ .}}
Under spectral flow, this becomes
\eqn\spfab{\eqalign{
 & \bra V_{J,{\bf J}'_R,J,{\bf J}'_L}^w (z_1,\zbar_1) V_{J,{\bf J}_R,J,{\bf
 J}_L}^{-w}
 (z_2,\zbar_2) \ket =
z_{12}^{-2\Delta_{J}} \zbar_{12}^{-2\bar{\Delta}_{J}}~\delta^2(J'_R+J_R)~\times \cr
 & \left( \delta(j+j'-1)
 + \delta (j-j') {\pi (2j-1) B(j) \Gamma (1-2j)\Gamma (j-iJ_R)\Gamma
 (j+iJ_L)\over
 \Gamma (2j)\Gamma (1-j-iJ_R)\Gamma (1-j+iJ_L)} \right). }}
with
\eqn\spfac{
J_R = {\bf J}_R + k\pi Tn\,, \quad J_L = {\bf J}_L -k\pi \bar{T}n }
and
\eqn\spfad{\eqalign{\Delta_{J} & =  \Delta_j - 2\pi T n {\bf J}_R- k
(\pi T n)^2  \cr
 \bar{\Delta}_{J}& =  \Delta_j + 2\pi \bar{T}n {\bf J}_L  -k
(\pi \bar{T} n)^2 \ . }}
Note again that in the Euclidean section the parameters are
complex valued. Therefore, the conformal weights $\Delta_J,\bar{\Delta}_J$
appear to be complex. However, we are interested in the 2-point
functions in the Lorentzian section, so we must again remember to
analytically continue all the parameters. Upon the analytic
continuation, the parameters are replaced by their values in the
Lorentzian BTZ geometry, which all take real values. In
particular, the conformal weights are again real.

As before, specific choices for the spin $j$ yield 2-point
functions for short or long strings. In particular we can obtain two-point
functions for strings which wind around the black hole.


\newsec{Discussion}

In this work we have discussed the computation of supergravity and
string theory correlation functions in the background of the extended
BTZ black hole. These were related to correlation functions in the
dual CFT living on the disconnected boundary of the spacetime.
The organizing principle was to use an appropriate analytic continuation
from Euclidean signature.  We conclude this paper
with a discussion of some open questions related to our work.

\item{$\bullet$}
The main outstanding issue is probably that of backreaction.  There
are good reasons to expect large effects in the extended geometry
due to the presence of closed timelike curves, and these could
 invalidate perturbation theory. This was pointed out in the original
work \BanadosWN, and related issues have been the subject of recent
discussion \refs{\LiuFT,\LawrenceAJ,\FabingerKR,\HorowitzMW}.
  In principle, this issue
can be addressed by studying scattering processes.   If perturbation
theory is indeed breaking down, one can still hope to make some progress
due to the fact that there is a well-defined holographic dual theory living
on the boundary.
\vskip.1cm

\item{$\bullet$}
The correlation functions that we have discussed are those of
non-normalizable vertex operators, corresponding to inserting operators
in the boundary CFT.  These are the correlation functions which can
be deduced by analytic continuation from Euclidean signature.  On
the other hand, in Lorentzian signature there are also normalizable
vertex operators, and in fact it is these that transform in unitary
representations of SL(2,R) $\times$ SL(2,R).  Experiments performed
by physical observers  correspond to transition amplitudes
between normalizable states in the Hilbert space, and so in string
theory should correspond to correlation functions of normalizable vertex
operators.
The normalizable correlation functions are in principle determined by
the non-normalizable ones, in the same way as in field theory the
transition amplitudes between normalizable states are determined by
the vacuum correlation functions, but this is somewhat indirect.
On the other hand, computing these directly
apparently requires working in Lorentzian
signature, and dealing with the unboundedness of the corresponding
worldsheet theory.
\vskip.1cm

\item{$\bullet$}
 One of the main motivations to study black hole spacetimes
in string theory is to address the information paradox.   To formulate the
paradox one really needs to consider a situation in which a black hole
forms from collapse and  then
evaporates completely (or conceivably  leaves a remnant),
since it is in this setup that unitarity and locality seem to clash.
In the eternal black hole there is no paradox: information thrown at
the black hole by one asymptotic observer may or may not be radiated
back to this observer; but if not, unitarity is retained simply by taking
into account the other regions into which the information can flow.
Studying the collapse scenario in string theory requires some new
ingredients.  In order to use worldsheet methods one needs a
conformal field theory representing matter collapsing to form a black
hole.  This is clearly a situation in which continuation to Euclidean
signature is impossible, so one has to learn to compute in Lorentzian
signature.   Hopefully, the lessons learned from studying toy models
like time-dependent orbifolds will be of use here.

\bigskip\medskip\noindent
{\bf Acknowledgements:} We thank Danny Birmingham and Ivo Sachs
for discussions, and Stephen Hwang and Yuji Satoh for additional
correspondence. S.H. was supported in part by the Magnus Ehrnrooth
Foundation, E.K-V. was suppported in part by the Academy of
Finland,  and P.K. was supported by NSF grant PHY-0099590. E.K-V.
would also like to thank the University of California at Los
Angeles and Institute for Theoretical Physics in Amsterdam for
hospitality while this work was in progress.

\listrefs

\end